\documentclass[10pt]{article}  
\usepackage{amssymb}
\usepackage{amsthm}    
\usepackage{graphicx}          
\usepackage{amsmath}         
\usepackage{mathtools}
\usepackage{mathrsfs}
\mathtoolsset{showonlyrefs}		
\usepackage{bbm}
\usepackage{authblk}

\usepackage{multirow}
\usepackage[most]{tcolorbox} 
\usepackage{xcolor}          

\usepackage{caption}
\usepackage{subcaption}
\usepackage{pgfplotstable}  
\usepackage{booktabs}       
\usepackage{tcolorbox}
\usepackage{nicematrix}

\usepackage{makecell}

\usepackage{dsfont}
\usepackage{listings}
\usepackage{titlepic}
\usepackage{xcolor}
\usepackage{array}
\usepackage{booktabs}
\usepackage{bm}
\usepackage{enumitem}
\usepackage{mathalfa}
\usepackage{todonotes}
\usepackage{hyperref}
\usepackage{geometry}

\usepackage[ruled]{algorithm2e}

\usepackage{enumitem}
\usepackage{braket}
\usepackage{physics}

\usepackage{algorithmic}

\usepackage{listings}

\usepackage{subcaption}

\newcommand{\fockket}[1]{\left| #1 \rangle\!\rangle \right.}

\usepackage[
backend=bibtex,
style=numeric,
isbn=false,
maxnames=99, 
]{biblatex}

\DeclareSourcemap{
  \maps[datatype=bibtex, overwrite]{
    \map{
      \step[fieldset=address, null]
      \step[fieldset=editor, null]
      \step[fieldset=location, null]
    }
  }
}

\AtEveryBibitem{%
  	\clearfield{issn} 
  	\clearfield{doi} 
  	\clearfield{urldate}
	\clearfield{month}
	\clearfield{adress}
  	\ifentrytype{online}{}{
    \clearfield{url}
  }
}

\hypersetup{
    colorlinks=true,
    linkcolor=blue,
    filecolor=magenta,      
    urlcolor=cyan,
    citecolor = blue,
}

\theoremstyle{plain}
\newtheorem{thm}{Theorem}[section]
\newtheorem{theorem}{Theorem}[section]
\newtheorem{definition}[thm]{Definition}

\newtheorem{lemma}[thm]{Lemma}

\newtheorem{corollary}[thm]{Corollary}

\newtheorem{assumption}{Assumption}
\newtheorem{proposition}[thm]{Proposition}
\newtheorem{remark}[thm]{Remark}

\newtheorem{example}[thm]{Example}

\numberwithin{equation}{section}

\newcommand{\bxi}{\boldsymbol{\xi}}

\DeclareMathOperator*{\argmin}{arg\,min}
\newcommand{\rbracket}[1]{\left(#1\right)}      
\newcommand{\cbracket}[1]{\left\{#1\right\}}      

\newcommand{\innerp}[1]{\langle{#1}\rangle}

\def\mL{\mathcal{L}}

\def\mS{\mathcal{S}}
\def\mN{\mathcal{N}}

\def\mA{\mathcal{A}}
\def\mV{\mathcal{V}}
\def\mB{\mathcal{B}}
\def\mQ{\mathcal{Q}}
\def\mR{\mathcal{R}}

\def\mK{\mathcal{K}}

\def\C{\mathbb{C}}

\newcommand{\bP}{\mathbf{P}}
\newcommand{\bS}{\mathbf{S}}

\newcommand{\bG}{\mathbf{G}}
\newcommand{\bA}{\mathbf{A}}
\newcommand{\bb}{\mathbf{b}}

\newcommand{\bC}{\mathbf{C}}
\newcommand{\bD}{\mathbf{D}}

\newcommand{\bW}{\mathbf{W}}

\newcommand{\bL}{\mathbf{L}}
\newcommand{\bV}{\mathbf{V}}
\newcommand{\bU}{\mathbf{U}}
\newcommand{\bE}{\mathbf{E}}

\newcommand{\bX}{\mathbf{X}}
\newcommand{\bK}{\mathbf{K}}

\newcommand{\bJ}{\mathbf{J}}

\renewcommand{\vec}{\mathbf{vec}}
\newcommand{\mat}{\mathbf{mat}}

\usepackage{todonotes}

\def \id {{\rm id}}

\title{A Unified Blockwise Measurement Design for Learning Quantum Channels and Lindbladians via Low-Rank Matrix Sensing}
\author{Quanjun Lang} 
\author{Jianfeng Lu}
\affil{Duke University}
\date{}

\begin{document}

\maketitle

\begin{abstract}
	Quantum superoperator learning is a pivotal task in quantum information science, enabling accurate reconstruction of unknown quantum operations from measurement data. We propose a robust approach based on the matrix sensing techniques for quantum superoperator learning that
    extends beyond the positive semidefinite case, encompassing both quantum channels and Lindbladians. We first introduce a randomized measurement design using a near-optimal number of measurements. By leveraging the restricted isometry property (RIP), we provide theoretical guarantees for the identifiability and recovery of low-rank superoperators in the presence of noise. Additionally, we propose a blockwise measurement design that restricts the tomography to the sub-blocks,
    significantly enhancing performance while maintaining a comparable scale of measurements. We also provide a performance guarantee for this setup. Our approach employs alternating least squares (ALS) with acceleration for optimization in matrix sensing. Numerical experiments validate the efficiency and scalability of the proposed methods. 
\end{abstract}

\tableofcontents
\section{Introduction}

The characterization of quantum systems is a central problem in quantum information science, requiring accurate reconstruction of superoperators that describe system dynamics. Quantum channels, which are completely positive and trace-preserving (CPTP) maps, model discrete-time evolution in quantum systems. The task of reconstructing the quantum channels from experimental data is called \textit{quantum channel tomography}, or \textit{quantum process tomography} in the literature \cite{chuang1997prescription, mohseni2008quantum}. 

A key challenge in this reconstruction is the inefficient scaling with system size. For an unknown, generic quantum channel acting on an $N$-dimensional quantum system, approximately $M \sim N^4$ expectation values are required. On the other hand, inherent structures can be leveraged for many physically relevant processes. In this work, we assume the quantum channel has a low Kraus rank. By representing the quantum channel using a matrix basis and a reshaping linear transformation, we can describe it using a low-rank positive semidefinite matrix, known as the Choi matrix. 
Recent advances in compressed sensing and matrix sensing have facilitated the efficient recovery of quantum channels with low Kraus rank from limited measurements \cite{klieschGuaranteedRecoveryQuantum2019, baldwinQuantumProcessTomography2014, rodionov2014compressed, flammia2012quantum}, which extends approaches in quantum states tomography \cite{kueng2017low, liuUniversalLowrankMatrix2011}, as they share the positive semidefinite property. 

However, most prior approaches focus on positive semidefinite matrices, which do not account for the distinct spectral properties of Lindbladians. The Lindblad-Gorini-Kossakowski-Sudarshan quantum master equation (QME) describes the continuous-time evolution of open quantum system dynamics under Markovianity assumption \cite{lindblad1976generators, gorini1976completely}
\begin{equation*}
	 \frac{d}{dt} \rho = \mL \rho := -i[H, \rho] + \sum_{k = 1}^{N_J} (J_k \rho J_k^\dagger - \frac{1}{2} J_k^\dagger J_k \rho - \frac{1}{2} \rho J_k^\dagger J_k ), 
\end{equation*}
where $H$ is the Hamiltonian that captures the unitary evolution, and $J_k$ are jump operators that characterize the dissipative evolution due to the interactions with the environment. The superoperator $\mL$ is called \textit{Lindbladian}. While one can learn the Lindbladian via learning the channel $\mV = e^{t \mL}$ and then recover $\mL$, here we aim at a more direct approach.
Unlike quantum channels, Lindbladians exhibit negative eigenvalues in their Choi-equivalent matrix representation, presenting unique challenges for superoperator reconstruction. 

In this work, we propose a unified approach for learning Lindbladian and quantum channels using a low-rank matrix sensing framework. We aim to learn a superoperator of the following form
\begin{equation}
	\mK \rho= \sum_{k = 1}^{r_+} V_k\rho V_k^\dagger - \sum_{k = 1}^{r_-} U_k \rho U_k^\dagger,
\end{equation}
where $V_k$ and $U_k$ are operators in $\C^{N \times N}$, and $r_+$ and $r_-$ represent the number of positive and negative parts, corresponding to the number of positive and negative eigenvalues of the reshaped Choi matrix. We assume the rank $r = r_+ + r_-$ is small relative to $N^2$, which is the total dimension of $\mK$. Specifically, the problem corresponds to quantum channel tomography when $r_- = 0$, and $r_- = 1$ for Lindbladian learning. Our method accommodates the negative eigenvalues in Lindbladians by leveraging the restricted isometry property (RIP) to ensure theoretical performance guarantees. Optimization is performed via alternating least squares (ALS), a widely used non-convex algorithm in matrix sensing and tensor reconstruction.

To further simplify the optimization landscape, we adopt a first-row blockwise measurement design, which reduces the learning of the superoperator $\mK$ of size $N^2 \times N^2$ into its sub-blocks of sizes $N \times N$, therefore improves the computational efficiency. Moreover, we show that such a measurement design has a performance guarantee provided by the matrix sensing theory. Numerical experiments confirm the effectiveness of the proposed approach, aligning with theoretical predictions and demonstrating its practical utility in quantum process tomography and Lindbladian learning.

This paper is organized as follows: Section \ref{sec_preliminary} introduces the preliminary for quantum superoperators, vectorization and reshaping, matrix sensing theory, and RIP condition. Section \ref{sec_3_random_measurement} provides a direct reconstruction algorithm using random measurement design, with theoretical guarantees based on RIP conditions. Section \ref{sec_methods_all} further improves the previous method by introducing a first-row blockwise measurement design and corresponding algorithms. Section \ref{sec_Numerics} shows the numerical experiments that validate our proposed methods. We conclude with a discussion of potential extensions and applications to broader quantum systems in Section~\ref{sec_conclusion}.

\subsection{Related works}

\subsubsection{Matrix sensing}\label{sec_1_1_Matrix_sensing}
Matrix sensing is a generalization of compressed sensing \cite{candes2006stable, donoho2006compressed, davenport2012introduction}. In the matrix sensing problem, it is assumed that the unknown matrix $\bX \in \C^{d_1 \times d_2}$ has low-rank $r \ll \min(d_1, d_2)$. We use known sensing matrices $\bA_1, \dots, \bA_M \in \C^{d_1 \times d_2}$ and possibly noisy linear observations $\bb_m = \tr [\bA_m^\dagger \bX]$ to recover $\bX$. Such a problem appears in many applications, including multi-task learning \cite{obozinski2010joint}, synchronization \cite{singer2011angular, liu2023unified, wang2013exact}, graph-related dynamical systems \cite{mardani2012dynamic, lang2024interacting}, and quantum tomography \cite{gross2010quantum}.


In recent years there has been tremendous progress in understanding how to solve such problems. One can consider the convex surrogate of the low-rank constraint using nuclear norm minimization \cite{rechtGuaranteedMinimumRankSolutions2010, candes2012exact, kueng2017low, candes2012exact, candes2010power}. There are also methods based on singular values decomposition (SVD) \cite{jain2010guaranteed, cai2010singular}. Beyond convex optimization methods, non-convex methods have been also proposed and analyzed \cite{xu2023power, daubechies2010iteratively, blumensath2009iterative, liao2019adaptive, blumensath2010normalized, tseng2001convergence, kim2014algorithms}. See \cite{davenport2016overview} for a thorough overview. 


A fundamental concept in sensing theory is the restricted isometry property (RIP), which quantifies the condition number of the sensing operator when restricted to low-rank matrices (see Definition \ref{def_rip}). The feasibility and performance of matrix sensing algorithms are closely tied to the RIP constant \cite{geNoSpuriousLocal2017}. For example, it is known that sensing matrices with iid Gaussian entries satisfy the RIP condition \cite{rechtGuaranteedMinimumRankSolutions2010}. 


In this work, we adopt the alternating least squares (ALS) algorithm \cite{kroonenberg1980principal, jain2013low, zhong2015efficient}, which is one of the non-convex optimization methods \cite{xu2013block}. We decompose the matrix $\bX = UV^\dagger$, where $U \in \C^{d_1 \times r}$ and $V \in \C^{d_2 \times r}$. The algorithm alternates between updating $U$ and $V$ in each iteration. The convergence properties of the alternating least squares algorithm for rank-one matrix sensing have been rigorously analyzed in \cite{lee2023randomly}. However, the question of global convergence for higher-rank matrix sensing remains an open problem. Notably, global convergence results have been established for tensor approximation using ALS \cite{yang2023global}.

\subsubsection{Quantum process tomography}
Matrix sensing and compressed sensing tools are widely used in quantum state tomography \cite{gross2010quantum, shabani2011efficient, kimmel2017phase}. In particular, the RIP condition has been proven to hold for random Pauli observables \cite{liuUniversalLowrankMatrix2011} under an incoherent condition \cite{gross2011recovering}. 


Using the Choi-Jamiołkowski representation of a quantum channel, sensing techniques can be applied to quantum process tomography \cite{baldwinQuantumProcessTomography2014, rodionov2014compressed, klieschGuaranteedRecoveryQuantum2019}. For example, in \cite{shabani2011efficient}, the authors introduced a quantum process tomography protocol under the assumptions of element-wise sparse with respect to a known basis. Under the assumption of low-Kraus rank, the authors in \cite{baldwinQuantumProcessTomography2014} proposed a method that ensures the CPTP property of the quantum channel, and it was numerically investigated in \cite{rodionov2014compressed}. Furthermore, in \cite{klieschGuaranteedRecoveryQuantum2019}, the authors compared it with commonly used nuclear norm minimization methods and established the theory of guaranteed recovery under an optimal measurement size.

\subsubsection{Lindbladian learning}\label{sec_related_work_Lindbladian}
Lindbladian learning is essential in studying open quantum systems \cite{samach2022lindblad, dobrynin2024compressed, onorati2023fitting, howard2006quantum}. The major difficulty of learning the Lindbladian lies in its high dimension and non-positive definiteness. Due to the nature of observing the quantum states, it is also expensive to have an accurate estimation of the derivative. Much work has been devoted to this field, including using polynomial interpolation with a local assumption on the Lindbladian based on Lieb-Robinson bound \cite{stilck2024efficient}. One can also avoid this problem of estimating the derivative by learning from steady states \cite{bairey2020learning}. The above approaches are called gradient matching, which gives a direct two-step road map for estimating ODE in general. See \cite{brunelParameterEstimationODE2008} for a detailed discussion on parameter learning in ODE, together with the consistency and asymptotic behavior of the gradient matching method. However, in this work, we only consider the second step, which assumes a noisy estimation of the derivatives. After acquiring the derivative estimation of the observed trajectories, the problem lies in the framework of matrix sensing, enabling a unified study of learning Lindbladian and channel tomography. It is worth mentioning that one can also apply nonlinear least squares \cite{li2005parameter, brunel2015tracking, hooker2009forcing}, which avoids derivative estimation. That method simulates the trajectory based on current parameters using nonlinear schemes and optimizes based on a least square loss \cite{wang2024simulation}.

If one focuses on closed quantum system instead of open ones, the task becomes Hamiltonian learning, which has seen tremendous progress in recent years, see e.g., \cite{li2020hamiltonian, Yu2022, Che2021, stilck2024efficient, Zubida2021, Granade2012, Gu2022, Bairey2019, ma2024learning}.

\begin{table}[!t] 
{\small
\begin{center} 
\caption{\, Notations} \label{tab:notation}
\begin{tabular}{ l  l | l  l }
\toprule 
Notation & Description & Notation & Description \\
\hline
$N$ & Dimension of Hilbert space & $\mV$ & Quantum channel \\
$N_J$ & Number of jump operators in  $\mL$ & $\mL$ & Lindbladian superoperator \\
$M_O$ & Number of observable operators &$\mK$ & General superoperator \\
$M$ & Number of measurements &  &  \\
$r$ & Rank of reshaped matrix $\bK$ &  &  \\
\hline
$\vec: \C^{N\times N} \to \C^{N^2}$ & Vectorization & $\rho$ & Density matrix \\
$\mat: \mB(\C^{N\times N}) \to \C^{N^2\times N^2}$ & Induced matrix by $\vec$ & $O$ & Observable operator \\
$\mR: \C^{N^2 \times N^2} \to \C^{N^2 \times N^2}$ & Reshaping operation & $H$ & Hamiltonian operator \\
$\bK = \mR(\mat(\mK))$ & Reshaped matrix of $\mK$ & $J_k$ & Jump operators \\
\bottomrule	
\end{tabular}  
\end{center}
}
\end{table}

\section{Preliminary}\label{sec_preliminary}

We aim to recover a quantum superoperator mapping from $\C^{N\times N}$ to $\C^{N \times N}$
\begin{equation}\label{eq_K_expression}
	\mK \rho= \sum_{k = 1}^{r_+} V_k\rho V_k^\dagger - \sum_{k = 1}^{r_-} U_k \rho U_k^\dagger,
\end{equation}
where the set of matrices $\{V_k\}$ and $\{U_k\}$ are orthogonal with Hilbert-Schmidt inner product, i.e.
\begin{equation}
    \tr[V_k^\dagger V_l] = \tr[U_k^\dagger U_l] = \lambda_k\delta_{kl}, \quad \tr[V_k^\dagger U_l] = 0, 
\end{equation}
for all $k, l$ in appropraite range. Here $r_+, r_- > 0$ represents the number of complex matrices $V_k$, $U_k$ in the positive and negative parts. We denote $r = r_+ + r_-$ and assume that $r \ll N^2$, where $N^2$ is the dimension of $\C^{N \times N}.$
 Suppose $\rho$ is Hermitian, it is clear that $\mK \rho$ is also Hermitian. We say that $\mK$ preserves Hermiticity. This setup is capable of describing both the quantum channels and the Lindbladians. See Example \ref{ex_quantum_channel} and Example \ref{ex_Linabladian} later.

\subsection{Vectorization and reshaping}
We now introduce the vectorization of $\rho$ and reshaping of $\mK$.
\subsubsection*{Vectorization}
We first define the vectorization operator $\vec:\C^{N\times N} \to \C^{N^2\times 1}$, which maps an $N\times N$ matrix to a $N^2 \times 1$ vector in the column-first order.  Then 
we define the following induced operator $\mat: \mB(\C^{N\times N}) \to \C^{N^2\times N^2}$ such that for any $\mK \in \mB(\C^{N\times N})$ and $\rho \in \C^{N\times N}$, 
\begin{equation}
    \vec(\mK \rho) =  \mat(\mK) \ \vec(\rho).
\end{equation}
\begin{lemma} \label{lem_vec} The $\vec$ operator has the following properties.
\begin{enumerate}
	\item $\vec(AXB) = (B^\top \otimes A)\vec(X)$. 
	\item $\vec$ is an isometry between $\C^{N\times N}$ with Hilbert-Schmidt inner product, and $\C^{N^2\times 1}$ with Euclidean inner product:
	for any $A, B \in \C^{N \times N}$, 
\begin{equation}
    \tr[A^\dagger B] = \innerp{A, B}_{\C^{N\times N}} = \sum_{i, j = 1}^N \overline{A}_{ij}B_{ij} = \vec(A)^\dagger\vec(B),
\end{equation}
where the superscript \( \dagger \) denotes the conjugate transpose and the overline denotes the element-wise complex conjugate. 

\end{enumerate}	
\end{lemma}
The proof is elementary and hence omitted. Note that we specify the dimension of the matrices in the Frobenius inner product. 
Applying vectorization to \eqref{eq_K_expression}, we have
\begin{equation}\label{eq_mat_K}
	\mat(\mK) = \sum_{k = 1}^{r_+} \overline{V}_k \otimes V_k - \sum_{k = 1}^{r_-} \overline{U}_k \otimes U_k.
\end{equation}

%
\subsubsection*{Reshaping}
If matrix $A \in \C^{N^2\times N^2}$, let $A_{ij}$ represents the $(i, j)$-th block when decomposing $A$ in to $N\times N$ submatrices. We define the following rearrangement operator $\mR$, 
\begin{equation}\label{eq_another_mR_Choi}
	A = 
	\begin{bmatrix}
 	A_{11} \dots A_{1n}\\
 	\vdots \ddots \vdots\\
 	A_{n1} \dots A_{nn} 
 	\end{bmatrix}, 
 	\quad
	\mR(A) = 
	\begin{bmatrix}
		\widetilde A_1, \dots, \widetilde A_{n}	
	\end{bmatrix}
	,\quad 
	\widetilde A_j = 
	\begin{bmatrix}
		\vec(A_{1j}), \dots, \vec(A_{nj})	
	\end{bmatrix}.
\end{equation}
In other words, $\mR$ vectorizes the submatrices to columns and stacks them in a column-first order. 
\begin{lemma}\label{lem_reshape_operator}
    This rearrangement operator $\mR$ is linear and satisfies the following properties.
    \begin{enumerate}
        \item $\mR(B\otimes C) = \vec(C) \vec(B)^\top$, for any $B, C \in \C^{N \times N}$. 
        \item $\mR$ is an isometry on $\C^{N^2\times N^2}$ equipped with the Frobenius inner product. 
        \item $\mR$ is an involution, namely $\mR((\mR(A)) = A$ for any $A \in \C^{N^2 \times N^2}$. 
    \end{enumerate}
\end{lemma}



\begin{proof}
    1. Let $A = B \otimes C$. Decompose $A$ into submatrices $A_{ij} = b_{ij}C$, where we use lower case $b_{ij}$ to represent the $(i, j)$-th element of the matrix B. The block $A_{ij}$ is rearranged to $b_{ij}\vec(C)$, which is the $((j-1)N+i)$-th column of $\mR(A)$. This equals to the $((j-1)N+i)$-th column of $\vec(C) \vec(B)^\top$, since $b_{ij}$ is the $((j-1)N+i)$-th element of $\vec(B)$. See the following equation for an example of $N=2$.
    \begin{tcolorbox}
$$
B \otimes C = 
\begin{bNiceMatrix}[margin]
\CodeBefore
  \begin{tikzpicture}
  \fill[blue!40,rounded corners] (1-|1) rectangle (3-|3); 
  \fill[red!40,rounded corners] (1-|3) rectangle (3-|5);  
  \fill[green!40,rounded corners] (3-|1) rectangle (5-|3); 
  \fill[yellow!40,rounded corners] (3-|3) rectangle (5-|5); 
  \end{tikzpicture}
\Body
b_{11}c_{11} & b_{11}c_{12} & b_{12}c_{11} & b_{12}c_{12} \\
b_{11}c_{21} & b_{11}c_{22} & b_{12}c_{21} & b_{12}c_{22} \\
b_{21}c_{11} & b_{21}c_{12} & b_{22}c_{11} & b_{22}c_{12} \\
b_{21}c_{21} & b_{21}c_{22} & b_{22}c_{21} & b_{22}c_{22}
\end{bNiceMatrix}
\overset{\mR}{\longrightarrow}
\begin{bNiceMatrix}[margin]
\CodeBefore
  \begin{tikzpicture}
  \fill[blue!40,rounded corners] (1-|1) rectangle (5-|2); 
  \fill[green!40,rounded corners] (1-|2) rectangle (5-|3); 
  \fill[red!40,rounded corners] (1-|3) rectangle (5-|4);  
  \fill[yellow!40,rounded corners] (1-|4) rectangle (5-|5); 
  \end{tikzpicture}
\Body
b_{11}c_{11} & b_{21}c_{11} & b_{12}c_{11} & b_{22}c_{11} \\
b_{11}c_{21} & b_{21}c_{21} & b_{12}c_{21} & b_{22}c_{21} \\
b_{11}c_{12} & b_{21}c_{12} & b_{12}c_{12} & b_{22}c_{12} \\
b_{11}c_{22} & b_{21}c_{22} & b_{12}c_{22} & b_{22}c_{22}
\end{bNiceMatrix}
= \vec(C)\vec(B)^\top
$$
\end{tcolorbox}
    \indent 2. It is straightforward to see that $\mR$ preserves the Frobenius inner product, as $\mR$ is a rearrangement of the entries. \\
    \indent 3. Note that the $(A_{ij})_{kl}$, which is the $(k, l)$-th element of $A_{ij}$, is mapped to the $((l-1)N+k, (j-1)N +i)$-th element of $\mR(A)$. Meanwhile, $(A_{ij})_{kl}$ is the $((i-1)N + k, (j-1)N+l)$-th element of $A$. It is clear to see that $\mR$ flips the position of $l$ and $i$. Therefore, $\mR$ is an involution.
%
%
\end{proof}


\begin{lemma}\label{lem_bK_hermitian}
	The reshaped matrix	$\bK = \mR(\mat(\mK))$ is Hermitian. 
\end{lemma}

\begin{proof}
By \eqref{eq_mat_K} and the first property in Lemma \ref{lem_reshape_operator},
\begin{equation}\label{eq_bK}
    \bK  = \sum_{k = 1}^{r_+} \vec(V_k)\vec(V_k)^\dagger  - \sum_{k = 1}^{r_-} \vec(U_k)\vec(U_k)^\dagger = \bV \bV^\dagger - \bU \bU^\dagger,
\end{equation}
where the last equality follows from the orthogonality assumption with
\begin{equation}
    \bV := 
    \begin{bmatrix}
        \vec(V_1^\dagger), \dots, \vec(V_{r_+}^\dagger)
    \end{bmatrix}
    \in \C^{N^2 \times r_+}, \quad
    \bU := 
    \begin{bmatrix}
        \vec(U_1^\dagger), \dots, \vec(U_{r_-}^\dagger)
    \end{bmatrix}
    \in \C^{N^2 \times r_-}.
\end{equation}
It is clear that $\bK$ is Hermitian. 
\end{proof}
In the later discussion, we decompose $\bK$ into $N \times N$ submatrices in $\C^{N \times N}$ denoted by $\bK_{ij}$, where $i$ representing the row and $j$ represents the column indices.

%

\begin{remark}
Note that a slightly different definition of the rearrangement operators was used  in \cite{vanloanApproximationKroneckerProducts1993}, whereas the reshaping operator is defined as 
	\begin{equation}\label{eq_another_mR_Van_loan}
	\mR_1(A) = 
	\begin{bmatrix}
		\widetilde A_1\\ \vdots \\ \widetilde A_{n}	
	\end{bmatrix}
	,\quad 
	\widetilde A_j = 
	\begin{bmatrix}
		\vec(A_{1j})^\top \\ 
		\vdots \\
		\vec(A_{Nj})^\top	
	\end{bmatrix},
\end{equation}
which is a mixed order for column-first vectorization and row-first in the alignment of the blocks. Such an arrangement can make 
$$\mR_1(A \otimes B) = \vec(A) \vec(B)^\top.$$
However, it is not an involution unless combined with the matrix transpose. Thus, we prefer the currently proposed reshaping without such distortion.
\end{remark}

\begin{example}[Quantum channels]\label{ex_quantum_channel}
	The evolution of a quantum system is characterized by CPTP maps called quantum channels. By the Choi-Kraus Theorem \cite{choi1975completely}, any quantum channel $\mV$ can be expressed by the Kraus representation, 
\begin{equation}\label{eq_mK}
    \mV \rho = \sum_{k = 1}^{r} V_k  \rho V_k^\dagger
\end{equation}
with $V_k \in \C^{N\times N}$ such that 
\begin{equation}
    \sum_{k = 1}^{r} V_k^\dagger V_k = \mathbbm{1}_{\C^{N\times N}},
\end{equation}
where $\mathbbm{1}_{\C^{N\times N}}$ is the identity operator on $\C^{N \times N}$ and $\{V_k\}_{k = 1}^r$ are called the Kraus operators. The Kraus representation is unique up to a unitary transformation. Here $r$ is called the Kraus rank. 

We can express the operator $\mV$ using a matrix under a basis of complex matrices. Suppose $E_{ij} = \ketbra{i}{j}$, where $\bra{i}$ represent the $i$-th canonical basis of $\C^N$, and $\ket{i} = \bra{i}^\dagger$. Then $\{E_{ij}\}_{i, j = 1}^N$ consist  a basis of $\C^{N \times N}$ and the Kraus operators can be expressed as vectors $\fockket{V_k} := \vec(V_k)$ in the Fock-Liouville space. The Choi matrix is defined as 
\begin{equation}
	\mR(\mat(\mV)) = \sum_{k= 1}^r \vec(V_k)\vec(V_k)^\dagger,
\end{equation}
which is the case of $r_-= 0$ in \eqref{eq_bK}.
\end{example}

\begin{example}[Lindbladian superoperators]\label{ex_Linabladian}
    The quantum master equation is given by $\frac{d}{dt}\rho = \mL \rho$, where
    \begin{equation}\label{eq_QME_main}
    	\mL \rho =  -i[H, \rho] + \sum_{k = 1}^{N_J} (J_k \rho J_k^\dagger - \frac{1}{2} J_k^\dagger J_k \rho - \frac{1}{2} \rho J_k^\dagger J_k ),
    \end{equation}
    where $H$ is Hermitian and represents the system Hamiltonian, and $J_k$ are jump operators with $k = 1\dots, N_J$.  Denote $Q = -iH - \frac{1}{2}\sum_{k = 1}^{N_J} J_k^\dagger J_k$, the above equation can be written as 
    \begin{equation*}
    	\mL \rho = Q\rho + \rho Q^\dagger + \sum_{k = 1}^{N_J}J_k \rho J_k^\dagger.
    \end{equation*}
	Apply vectorization on both sides, 
    \begin{equation*}
    	\mat(\mL) \vec(\rho) = (I \otimes Q)\vec(\rho) + (\overline{Q} \otimes I) \vec(\rho)  + \sum_{k = 1}^{N_J} \rbracket{\overline{J}_k \otimes J_k}\vec(\rho).
    \end{equation*}
    Then apply the rearrangement $\mR$, 
    \begin{equation}
    	\bL := \mR(\mat(\mL)) = \vec(Q)\vec(I)^\dagger + \vec(I)\vec(Q)^\dagger + \sum_{k = 1}^{N_J}\vec(J_k)\vec(J_k)^\dagger.
    \end{equation}
    It is clear that $\sum_{k = 1}^{N_J}\vec(J_k)\vec(J_k)^\dagger$ is positive semidefinite. Moreover, 
    \begin{equation*}
    	\vec(Q)\vec(I)^\dagger + \vec(I)\vec(Q)^\dagger = uu^\dagger - vv^\dagger,\quad u = \frac{\vec(I) + \vec(Q)}{\sqrt{2}}, \quad v = \frac{\vec(I) - \vec(Q)}{\sqrt{2}},
    \end{equation*}
    so that the matrix $\vec(Q)\vec(I)^\dagger + \vec(I)\vec(Q)^\dagger$ has one positive and one negative eigenvalues. Suppose $\cbracket{J_k}_{k = 1}^{N_J}$ are not degenerate in Hilbert-Schmidt inner product, we showed that $\bL$ has an expression as \eqref{eq_K_expression}, where $r_+ = N_J + 1$ and $r_- = 1$, therefore $\rank(\bL) = N_J +2$. 
    \end{example}

\subsection{Observables}
Given an input quantum state $\rho_0$, denote the output of $\mK$ as 
\begin{equation}
	\rho_1 = \mK \rho_0.
\end{equation}
Note that $\rho_1$ may not be a valid quantum state because we did not assume that $\mK$ is CPTP. However, $\rho_1$ remains Hermitian since $\mK$ preserves Hermiticity. Despite this, $\rho_1$ cannot be directly observed in practice. Any measurement involving $\rho_1$ is performed through its inner product with a Hermitian observable operator $O \in \mB(\C^N)$, and therefore the measurement $\innerp{\rho_1, O}_{\C^{N\times N}}$ is a real number. If $\mK$ is a quantum channel, such a measurement can be directly obtained. However, when $\mK$ represents a Lindbladian superoperator, $\rho_1$ corresponds to the time derivative of the evolving quantum state, meaning the measurement reflects the time derivative of the observable process. See Section \ref{sec_related_work_Lindbladian} for more discussions.



\begin{proposition}\label{prop_measurement_tensor_product}
    The Hilbert-Schmidt inner product of $\rho_1$ and $O$ can be written as 
    \begin{equation}\label{eq_rho1_K_equivalent_form}
        \innerp{\rho_1, O}_{\C^{N\times N}} = \innerp{\bK, \overline{\rho}_0 \otimes O}_{\C^{N^2\times N^2}} = \innerp{\rho_0^\top\otimes O^\dagger, \bK^\dagger}_{\C^{N^2\times N^2}}.
    \end{equation}
    In particular, if $\rho_0$, $O$ and $\bK$ are Hermitian, 
    \begin{equation}\label{eq_rho1_K_equivalent_form_Herm}
        \innerp{\rho_1, O}_{\C^{N\times N}} =  \innerp{\overline{\rho}_0 \otimes O, \bK}_{\C^{N^2\times N^2}}.
    \end{equation}
\end{proposition}

\begin{proof}

By the isometry property of $\vec$ in Lemma \ref{lem_vec},
\begin{equation}
    \innerp{\rho_1, O}_{\C^{N\times N}} = \innerp{\mK \rho_0, O}_{\C^{N\times N}} = \vec(\mK \rho_0)^\dagger \vec(O) = \vec(\rho_0)^\dagger \mat(\mK)^\dagger \vec(O),
\end{equation}
where the last term is a constant and equals its trace. Using the cyclic property of trace, 
\begin{equation}
    \vec(\rho_0)^\dagger \mat(\mK)^\dagger \vec(O) 
    = \tr[\mat(\mK)^\dagger \vec(O) \vec(\rho_0)^\dagger]= \innerp{\mat(\mK), \vec(O) \vec(\rho_0)^\dagger}_{\C^{N^2\times N^2}}.
\end{equation}
Recall that the reshaping operator $\mR$ is an isometry on $\C^{N^2 \times N^2}$, so that
\begin{equation}
    \innerp{\mat(\mK), \vec(O) \vec(\rho_0)^\dagger}_{\C^{N^2\times N^2}}= \innerp{\mR(\mat(\mK)), \mR(\vec(O) \vec(\rho_0)^\dagger)}_{\C^{N^2\times N^2}},
\end{equation}
whereas $\bK = \mR(\mat(\mK))$ by \eqref{eq_bK}. Moreover, we have by Lemma \ref{lem_reshape_operator}
\begin{equation}
	\vec(O)\vec(\rho_0)^\dagger = \mR(\overline{\rho}_0 \otimes O).
\end{equation}
Using the involution property of $\mR$, 
\begin{equation}
    \mR(\vec(O)\vec(\rho_0)^\dagger) = \mR(\mR(\overline{\rho}_0 \otimes O)) = \overline{\rho}_0 \otimes O.
\end{equation}
In the end, we have 
\begin{equation}
	\innerp{\rho_1, O}_{\C^{N\times N}} = \innerp{\bK, \overline{\rho}_0 \otimes O}_{\C^{N^2\times N^2}} = \innerp{\rho_0^\top\otimes O^\dagger, \bK^\dagger}_{\C^{N^2\times N^2}}.
\end{equation}
And \eqref{eq_rho1_K_equivalent_form_Herm} holds naturally if $\rho_0$, $O$ and $\bK$ are Hermitian.
\end{proof}

This enables us to formulate the problem using matrix sensing. See Section \ref{sec_matrix_sensing} and Section \ref{sec_Pauli_RIP}. Suppose we can choose $\rho_0$ to be the matrix $E_{ij}$, whose entries are all zeros except for its $(i, j)$-the entry equals to $1$, then we can isolate the tomography of $\bK$ into its subblock. However, since $E_{ij}$ are not valid density matrices, we have to introduce more initial states for an equivalent evaluation. We will discuss details in Section \ref{sec_one_row}.

\subsection{Matrix sensing}\label{sec_matrix_sensing}

We now introduce the preliminary of matrix sensing.
Suppose $\bX^\star \in \C^{d_1 \times d_2}$ is an unknown matrix of rank $r$. For a group of sensing matrices $\bA_m \in \C^{d_1 \times d_2}, m = 1,\dots, M$, we define the sensing operator,
\begin{equation}\label{eq_A_b_general}
	\mA(\bX) = \frac{1}{\sqrt{M}}\begin{bmatrix}
		\tr[\bA_1^\dagger \bX] \\ 
		\vdots \\
		\tr[\bA_M^\dagger \bX] 
	\end{bmatrix}
	\in \C^{M}.
\end{equation} 
Let $\bb^\star = \mA(\bX^\star)$, we assume the observation is given by $\bb = \bb^\star + \bxi$, where the entries of $\bxi$ are iid Gaussian random noise with standard deviation $\sigma$. To find $\bX^\star$, we need to solve the following nonconvex optimization problem, 
\begin{equation}\label{eq_loss_ALS_general}
	\argmin_{\bX \in \C^{d_1\times d_2}, \rank(\bX) = r} f(\bX) = \frac{1}{2}\norm{\mA(\bX) - \bb}^2 = \frac{1}{2M}\sum_{m = 1}^M \abs{\tr[\bA_m^\dagger \bX] - \bb_m}^2.
\end{equation}
It is natural to transform the problem to $\bX = UV^\dagger$, where $U \in \C^{d_1\times r}$ and $V \in \C^{d_2 \times r}$. With an abuse of notation, we also denote $f(U, V) = f(\bX)$. 
However, for the asymmetric case where $d_1 \neq d_2$, or $\bX^\star$ is not positive semidefinite, it is necessary to consider the regularizer,
\begin{equation}\label{eq_loss_ALS_general_reg}
	\argmin_{\bX \in \C^{d_1\times d_2}, \rank(\bX) = r} \widetilde{f}(\bX) = \frac{1}{2M}\sum_{i = 1}^M \abs{\tr[\bA_i^\dagger \bX] - \bb_i}^2 + \lambda \norm{U^\dagger U - V^\dagger V}_{\C^{r \times r}}^2
\end{equation}
with the regularization parameter $\lambda = 1/8$ proposed in \cite{geNoSpuriousLocal2017}. The regularization term in \eqref{eq_loss_ALS_general_reg} has been utilized in prior works, such as \cite{zheng2016convergence, park2017non}. Its purpose is to transform the asymmetric case into a symmetric one by imposing an additional constraint. 

To establish the theoretical foundation for analyzing the optimization landscape of the proposed problem, we first introduce the concept of a local minimum. 
\begin{definition}[Local minimum]
Suppose $\bX$ is a \textbf{local minimum} of $f(\bX)$, then we have
\[
\nabla f(\bX) = 0, \quad \nabla^2 f(\bX) \succeq 0.
\]
\end{definition}
Following this, we define the restricted isometry property (RIP), which describes the geometric property of the sensing operator over low-rank matrices.
\begin{definition}[RIP]\label{def_rip}
The sensing operator $\mA$  satisfies the $(r, \delta)$-RIP condition, if for every matrix $\bX$ of rank at most $r$,
\begin{equation}\label{eq_RIP}
	(1 - \delta)\|\bX\|_{\C^{d_1\times d_2}}^2  \leq \norm{\mA(\bX)}_2^2 \leq (1 + \delta)\|\bX\|_{\C^{d_1\times d_2}}^2.
\end{equation}
\end{definition}
The landscapes of the above optimization problems have been studied in the literature, which depend crucially on the RIP condition. 
\begin{theorem}[Theorem 8, \cite{geNoSpuriousLocal2017}]\label{thm_landscape_noiseless}
Suppose the measurement operator $\mA$ defined in \eqref{eq_A_b_general} satisfies $(2r, \frac{1}{20})$-RIP. Then for the noiseless matrix sensing objective \eqref{eq_loss_ALS_general_reg} with $\lambda = 1/8$, we have:
\begin{enumerate}
    \item All local minima $(U, V)$ satisfy $UV^\dagger = \bX^\star$.
    \item The objective function $\widetilde{f}$ is $\left(\varepsilon, \frac{\sigma_r^\star}{5}, \frac{20\epsilon}{\sigma_r^\star}\right)$-strict saddle, where $\sigma_r^\star$ denotes the $r$-th singular value of $M^\star$. A function $g(\cdot)$ is said to be $(\theta, \gamma, \zeta)$-strict saddle if, for any $\bX$, at least one of the following hold: (a) $\|\nabla g(\bX)\| \geq \theta$; (b) $\lambda_{\min}(\nabla^2 g(\bX)) \leq -\gamma$; or (c) $\bX$ is $\zeta$-close to the set of local minima.
\end{enumerate}
\end{theorem}

%
%
%

For the noisy case, we apply a subsequent result. 
\begin{theorem}[Theorem 32, \cite{geNoSpuriousLocal2017}]\label{thm_landscape_noisy}
Suppose the sensing operator $\mA$ defined in \eqref{eq_A_b_general} satisfies $(2r, \frac{1}{20})$-\textit{RIP}. Then for the noisy matrix sensing objective \eqref{eq_loss_ALS_general_reg} all local minima must satisfy 
\[
    \| U V^\dagger - \mathbf{X}^\star \|_{\C^{d_1\times d_2}} \leq O\left(\sigma \sqrt{\frac{r(d_1+d_2) \log M}{M}}\right)
\]
with high probability. 
\end{theorem}

We will rely on these two theorems as the primary tools for establishing the theoretical guarantees.  The original theorems are valid for real-valued matrices. We argue that the same proof holds for the complex case. 

\begin{remark}\label{rmk_delta_c}
It is important to notice that the magnitude of the sensing operator does not affect the RIP constant, and what matters is the condition number of the operator over the low-rank matrices. Suppose $\widehat{\mA}$ is a sensing operator with matrices $\widehat{A}_1, \dots, \widehat{A}_M$ and $\widehat{\bb}$ is the measurement results, so that for some positive constants $c_0$ and $c_1$,
\begin{equation}\label{eq_c0_c1}
	c_0\|\bX\|_{\C^{d_1\times d_2}}^2  \leq \norm{\widehat\mA(\bX)}_2^2 \leq c_1\|\bX\|_{\C^{d_1\times d_2}}^2.
\end{equation} 
It is equivalent to have $\widehat{\mA} = c\mA$, $\widehat{\bb} = c \bb$,  where $\mA$ satisfies the $(r, \delta)$-RIP condition. Here $c$ and $\delta$ are given by
	\begin{equation}\label{eq_delta_c}
		\begin{cases}
			c(1-\delta) = c_0\\
			c(1+\delta) = c_1
		\end{cases}
		\Rightarrow \quad
		\begin{cases}
			\delta = \frac{c_1 - c_0}{c_1 + c_0}\\
			c = \frac{c_0 + c_1}{2}
		\end{cases}.
	\end{equation}
	Using this setting, we need to solve the following optimization problem,
	\begin{equation}\label{eq_loss_scalling}
		\argmin_{\bX \in \C^{d_1\times d_2}, \rank(\bX) = r} \frac{1}{2}\norm{c\mA(\bX) - c\bb}^2.
	\end{equation}
		which is just a scalar multiplication of \eqref{eq_loss_ALS_general}. Therefore the same results in Theorem \ref{thm_landscape_noiseless} holds. In the noisy case, suppose $\widehat\bb$ is polluted by iid Gaussian noise with standard deviation $\widehat{\sigma}$, it is equivalent to the noisy observation problem with $\mA$, $\bb$ and noise standard deviation $\sigma = \widehat{\sigma}/c$. 
\end{remark}


\subsection{Restricted isometry property for Pauli measurements}\label{sec_Pauli_RIP}
We now examine the conditions under which the sensing operator satisfies the RIP. It is well-established that the RIP condition holds for a sufficient number of sensing matrices with iid Gaussian entries \cite{candes2012exact}. However, this result is not directly applicable to quantum tomography, as the sensing matrices must satisfy additional structural constraints. In \cite{liuUniversalLowrankMatrix2011}, for the case where $d:= d_1 = d_2$, the author demonstrated that the RIP condition holds when the sensing matrices are incoherent with respect to the canonical basis of $\C^{d \times d}$. A similar condition appears in \cite{gross2011recovering}, which essentially requires that the sensing matrices should not be overly aligned with the standard basis vectors. See \cite{candes2012exact} for a detailed study of incoherence. 

\begin{definition}[Incoherence]\label{def_incoherent_matrix}
	Suppose an orthonormal basis $\{W_1, \dots W_{d^2}\}$ of $\C^{d \times d}$ with respect to the Hilbert-Schmidt inner product. Then we say that it is \textbf{incoherent} if the $W_i$ all have small operator norm:
	\begin{equation}\label{eq_incoherence}
		\norm{W_i} \leq K/\sqrt{d}, 
	\end{equation}
	where $K$ is a constant. 
\end{definition}
\begin{example}\label{ex_scaled_Pauli}
	Pauli matrices are a class of convenient quantum observables. For system of $n$ qubits with Hilbert space dimension $d = 2^n$, Pauli matrices are of the form $P_1 \otimes \dots \otimes P_n$, where each $P_i$ is a $2\times 2$ matrix chosen from the following four possibilities:
	\begin{equation}
		I = 
		\begin{pmatrix}
 			1 & 0\\
 			0 & 1
		\end{pmatrix}
		,\quad
		\sigma_x = 
		\begin{pmatrix}
 			0 & 1\\
 			1 & 0
		\end{pmatrix}
		,\quad
		\sigma_y = 
		\begin{pmatrix}
 			0 & -i\\
 			i & 0
		\end{pmatrix}
		,\quad
		\sigma_z = 
		\begin{pmatrix}
 			1 & 0\\
 			0 & -1
		\end{pmatrix}.
	\end{equation}
	In particular, the scaled Pauli matrices $(P_1\otimes \cdots \otimes P_n)/\sqrt{d}$ are orthonormal basis with $K = 1$. 
\end{example}


We recall the sufficient sample size for randomly selected Pauli matrices to achieve RIP property, which has been studied extensively in \cite{liuUniversalLowrankMatrix2011}. While the original result has slightly different definitions for the sensing operator $\mA$ and the RIP condition, we have adapted it to align with our framework for consistency.

\begin{theorem}[Theorem 2.1 in \cite{liuUniversalLowrankMatrix2011}]\label{thm_Pauli_RIP}
	Let $\{W_1, \dots, W_{d^2}\}$ be an orthonormal basis for $\mathbb{C}^{d \times d}$ that is incoherent in the sense of \eqref{eq_incoherence}. Suppose we independently and uniformly choose the sensing matrices $\bA_1, \dots, \bA_M$ from $\{W_1, \dots, W_{d^2}\}$. Then, for $M = C K^2 r d \log^6 d$, where $C= O(1/\widehat\delta^2)$ is a constant, the following inequality holds with high probability over the selection of the sensing matrices,
    \begin{equation}\label{eq_RIP_Pauli_different}
        \frac{1-\widehat\delta}{d}\norm{X}_{\C^{d \times d}} \leq \norm{\mA(X)}_2 \leq \frac{1+\widehat\delta}{d}\norm{X}_{\C^{d \times d}}.
    \end{equation}
    In particular, if the basis matrices are scaled Pauli as defined in Example \ref{ex_scaled_Pauli}, $K = 1$. 
\end{theorem}
Using Remark \ref{rmk_delta_c}, we can derive 
\begin{equation}\label{eq_delta_c_RIP}
	c = \frac{1 + \widehat\delta^2}{d^2} = O(d^{-2}), \quad {\delta} = \frac{2\widehat\delta}{1 + \widehat\delta^2} = O(\widehat\delta),
\end{equation}
where the big-O notation represents the order when $\widehat\delta \to 0$ and $d \to \infty$. We can now connect the results in \cite{liuUniversalLowrankMatrix2011} and \cite{geNoSpuriousLocal2017}. 

\begin{theorem}\label{thm_incoherence_RIP_landescape}
	Suppose we independently and uniformly choose $\bA_1, \dots, \bA_M$ from scaled Pauli matrices. Then, for $M \geq C r d \log^6 d$ where $C= O(1/\delta^2)$ is a constant, $\frac{1}{c}\mA$ satisfies the $(r, \delta)$-RIP condition with high probability over the selection of the matrices, with $c$ and $\delta$ defined in \eqref{eq_delta_c_RIP}. Moreover, if $\delta \leq \frac{1}{20}$, the following hold for the matrix sensing objective \eqref{eq_loss_ALS_general_reg} with $\lambda = c/8$.
	\begin{enumerate}
		\item For the noiseless case $(\sigma = 0)$, all local minima $(U, V)$ satisfy $UV^\dagger  = \bX^\star$. 
		\item For the noisy case $(\sigma > 0)$, all local minima must satisfy 
	\[
	    \| U V^\dagger - \mathbf{X}^\star \|_{\C^{d\times d}} \leq O\left(\sigma d^2\sqrt{\frac{rd\log M}{M}}\right)
	\]	
	with high probability.
	\end{enumerate}
\end{theorem}


\subsection{Alternating least squares}
Consider the loss function in \eqref{eq_loss_ALS_general} and let $\bX = UV^\dagger$, where $U \in \C^{d_1 \times r}$ and $V \in \C^{d_2 \times r}$. The loss function is quadratic when one of $U$ or $V$ is fixed, which suggests an alternating update of $U$ and $V$ using least squares (ALS), see Algorithm \ref{alg_ALS_N}. We refer back to Section \ref{sec_1_1_Matrix_sensing} for related works. The ALS is known to have local convergence and global convergence in tensor completion \cite{yang2023global, xu2013block}. Its global convergence in matrix sensing problems has been justified for rank-1 matrices with random initialization \cite{lee2023randomly}. However, the global convergence of ALS in matrix sensing with higher rank ($r \geq 2$) is still an open problem.

\begin{algorithm}
\caption{Alternating least squares (ALS)}\label{alg_ALS}
    \hspace*{\algorithmicindent} 
    \textbf{Input:} Sensing operator $\mA$, measurements $\bb$, rank $r$, maximal iteration number $n_{maxiter}$ and minimum discrepancy per iteration $\gamma$.\\
    \hspace*{\algorithmicindent} 
    \textbf{Output:} Estimated low-rank matrix $\bX$.\\
\begin{algorithmic}[1]
\STATE{Randomly initialize $U_0 \in \C^{d_1 \times r}$.}
\FOR{$\tau = 1, \dots, n_{maxiter}$} 
    \STATE{Estimate the matrix $V_\tau$ by solving $\argmin_{V \in \C^{d_2 \times r}} f (U_{\tau-1}, V)$ using least squares.}
    \STATE{Estimate the matrix $U_\tau$ by solving $\argmin_{U \in \C^{d_1 \times r}} f (U, V_{\tau})$ using least squares.}
    \STATE{Exit loop if $\|U_\tau - U_{\tau-1}\| \leq \gamma \|U_{\tau-1}\|$ and $\|V_\tau - V_{\tau-1}\| \leq \gamma \|V_{\tau-1}\|$.}
\ENDFOR
\RETURN{$\bX = U_\tau V_\tau^\dagger$}
\end{algorithmic}
\end{algorithm}

\begin{remark}
	The complexity of the ALS depends on the number of iterations and the complexity of linear regressions. The linear regressions can be solved explicitly with complexity $O(r^2d^2M)$, which is due to the complexity of SVD where $d= d_1 + d_2$. Using gradient descent (GD) or stochastic gradient descent (SGD), this complexity can be improved to $O(rdMT)$ or $O(rdT)$, where $T$ is the number of iterations. With a certain level of accuracy $\varepsilon$, the number of iterations for GD is $O({\kappa \log(1/\varepsilon)})$, where $\kappa$ is the condition number of the normal matrix of the linear regression. For SGD, since the objective function is strongly convex, we have $T\sim O(1/\varepsilon)$ with an optimal choice of step size.
	
	However, determining the total number of iterations required for ALS is a more complex challenge. In \cite{lee2023randomly}, the authors showed the case of $r = 1$ and iid Gaussian sensing matrices, 
    the iterations needed to reach $\varepsilon$ accuracy is $O(\frac{\log d}{\log \log d} + \frac{\log(1/\varepsilon)}{\log\log d})$ with small falling probability.
\end{remark}

To further enhance the convergence speed, we apply the Nesterov acceleration in ALS \cite{mitchell2020nesterov}. Similar acceleration techniques can be found in \cite{peng2024block, guminov2021combination}. The idea is to update $U, V$ based on their history, similar to the Nesterov acceleration in gradient descent, which computes the gradient based on a shifted position using momentum. We apply the notation in \cite{mitchell2020nesterov}. Let $\bX_\tau = U_\tau V_\tau^\dagger$ and denote the update of $U$ and $V$ using one step of ALS as $\bX_{\tau + 1} = q_{ALS}(\bX_\tau)$. The Nesterov method suggests an update where $\bX_{\tau + 1} = q_{ALS}(\bX_\tau + \beta_\tau(\bX_\tau - \bX_{\tau-1}))$. If the loss function increases so that $f(\bX_\tau) \geq \eta f(\bX_{\tau -1})$, the algorithm restarts by setting $\bX_\tau = \bX_{\tau-1}$, which gets back to the original ALS update steps. We apply $\beta_\tau = 1$ and $\eta = 1.2$ suggested in \cite{mitchell2020nesterov} and conclude this method in Algorithm \ref{alg_ALS_N}.


\begin{algorithm}
\caption{Nesterov-ALS \cite[Algorithm 1]{mitchell2020nesterov}}\label{alg_ALS_N}
    \hspace*{\algorithmicindent} 
    \textbf{Input:} Sensing operator $\mA$, measurements $\bb$, rank $r$, maximal iteration number $n_{maxiter}$ and minimum discrepancy per iteration $\gamma$, restart parameter $\eta = 1.2$.\\
    \hspace*{\algorithmicindent} 
    \textbf{Output:} Estimated low-rank matrix $\bX$.\\
\begin{algorithmic}[1]
\STATE{Randomly initialize $\bX_0, \bX_1$.}
\FOR{$\tau = 2, \dots, n_{maxiter}$} 
    \IF{$f(x_\tau) \geq \eta f(x_{\tau -1})$}
        \STATE{$\bX_\tau = \bX_{\tau - 1}, \beta = 0$.}
    \ELSE
        \STATE{$\beta = 1$.}
    \ENDIF
    \STATE{Exit loop if $\|\bX_\tau - \bX_{\tau-1}\| \leq \gamma \|\bX_{\tau-1}\|$.}
    \STATE{$\bX_{\tau + 1} = q_{ALS}(\bX_\tau + \beta(\bX_\tau - \bX_{\tau -1}))$.}
\ENDFOR
\RETURN{$\bX = \bX_{\tau+1}$}
\end{algorithmic}
\end{algorithm}

\begin{remark}\label{rmk_ALS_not_reg}
	Note that the ALS algorithm does not comply with the regularized objective function \eqref{eq_loss_ALS_general_reg} since the regularization introduces non-quadratic behavior in the objective function when either $U$ or $V$ is fixed. In our proposed algorithm for optimization, we omit this additional regularization term. Empirical evidence indicates that this omission does not impact the results.
\end{remark}

\section{Measurement design and algorithms}\label{sec_methods_all}
We aim to recover the superoperator $\mK$ of the form \eqref{eq_K_expression}, assuming that the rank $r$ is known along with an ideal number of given initial states, observables, and measurement outputs.



\subsection{Random measurement design}\label{sec_3_random_measurement}
Our first measurement design uses observation pairs $\{\rho_0^m, O^m\}_{m = 1}^M$. Recall from Proposition \ref{prop_measurement_tensor_product}, for each given pair of initial state and observable, the true measurement data is given by
\begin{equation}\label{eq_b_noise}
    b_0^{\star m} = \tr\left[\mK^\star(\rho_0^m)^\dagger O^m\right] = \tr\left[(\overline{\rho}_0^m \otimes O^m )^\dagger {\bK^\star} \right], 
\end{equation} 
where the superscript $\star$ represents the true superoperator or true reshaped matrix. With the presence of observation noise, the problem lies in the matrix sensing framework with
\begin{align}\label{eq_A_b_all}
    \mA_0 (\bK) = 
    \begin{bmatrix}
	\tr[\bA_1^\dagger\bK ]\\
	\vdots \\ 
	\tr[\bA_M^\dagger \bK  ]\\
	\end{bmatrix}
        , 
	\quad 
	\bb_0 = \bb^\star_0  + \bxi \in \C^{M},
\end{align}
where $\bA_m = \overline{\rho}_0^m \otimes O^m$, $\bb_0^\star = [b^{\star 1}_0, \dots, b^{\star M}_0]^\top$ and $\bxi = [\xi^1, \dots, \xi^M]^\top$ with $\xi^m \sim \mN(0, \sigma^2)$. We aim to find the minimizer $\bK$ of a quadratic loss,
\begin{equation}\label{eq_loss_all}
    \quad \argmin_{\bK \in \C^{N^2 \times N^2},\ \rank(\bK) = r}f_0(\bK) = \frac{1}{2M}\norm{\mA_0(\bK) - \bb_0}^2.\
    \tag{$\mQ_{0}$}
\end{equation}
Note that this is the same objective as \eqref{eq_loss_ALS_general} with dimension $d = N^2$. However, it is necessary to introduce the regularization as \eqref{eq_loss_ALS_general} since $\bK$ is not assumed to be positive semidefinite, as we want to include the Lindbladian learning. The theoretical guarantee is provided based on \eqref{eq_loss_ALS_general}, and we apply ALS for optimization.

By Theorem \ref{thm_Pauli_RIP}, the incoherence of $\bA_m$ plays a crucial role in efficiently achieving a small RIP constant. Recall from Example \ref{ex_scaled_Pauli}, we can take 
\begin{equation}\label{eq_O_rho_Pauli}
	\overline{\rho}_0^m \otimes O^m = (\overline{P}_1 \otimes \cdots \otimes \overline{P}_n \otimes  P_{n+1} \cdots \otimes  P_{2n} )/N,
\end{equation}
where $n$ is the number of qubits with $N = 2^n$, and $P_i$ are uniformly iid samples of the four $2 \times 2$ Pauli matrices. It is important to note that Pauli matrices are not valid density matrices, as they generally do not have trace one. This issue will be addressed later in Remark \ref{rmk_Pauli_eigen}. The matrices in \eqref{eq_O_rho_Pauli} form an orthonormal basis of $\C^{N^2\times N^2}$ and are incoherent with constant $K = 1$. Applying the Theorem \ref{thm_incoherence_RIP_landescape} with $d = N^2$, we have the following result. 
\begin{corollary}\label{cor_ALS_N_2}
	Suppose initial states $\rho_0^m$ and observables $O^m$ are uniform iid samples of the scales Pauli matrices. Then for $M \geq C r N^2 \log^6 N$ where $C= O(1/\delta^2)$ is a constant, $\frac{1}{c}\mA_0$ satisfies the $(r, \delta)$-RIP condition with high probability over the selection of the matrices, with $c$ and $\delta$ defined in \eqref{eq_delta_c_RIP}. Moreover, if $\delta \leq \frac{1}{20}$, the following hold for the matrix sensing objective \eqref{eq_loss_ALS_general_reg} using $\mA_0$ and $\bb_0$ with $\lambda = c/8$. 
	\begin{enumerate}
		\item For the noiseless case $(\sigma = 0)$, all local minima $(U, V)$ satisfies $UV^\dagger  = \bK^\star$. 
		\item For the noisy case $(\sigma > 0)$, all local minima must satisfy 
	\[
	    \| U V^\dagger - \mathbf{K}^\star \|_{\C^{N^2\times N^2}} \leq O\left(\sigma N^5\sqrt{\frac{r\log M}{M}}\right)
	\]	
	with high probability.
	\end{enumerate}
\end{corollary}
Note that this result is worse than \cite{klieschGuaranteedRecoveryQuantum2019}, which requires $M = CrN^2\log N$ measurements. The extra power of the $\log$ is due to the absence of the positive definiteness assumption of the matrix $\bK$. 


\begin{remark}\label{rmk_Pauli_eigen}
To equivalently evaluate $\mK$ over a scaled Pauli matrix $P$, we first find its eigenvalues and eigenvectors $\lambda_j$ and $\ket{\phi_j}$. Then for any observable $O$, the following holds,
\begin{equation}
    \tr[\mK(P)O]
	 = \sum_{j = 1}^N \lambda_j \tr[O \mK(\ketbra{\phi_j})].
\end{equation}
To estimate this quantity, we repeat the following procedure and compute the average: uniformly select \( j \in [N] \), prepare the state \( \ket{\phi_j} \), apply the operator \(\mK\), measure using \( O \), and scale the result by \(\lambda_j\).  This approach was first mentioned in \cite{flammia2012quantum}, which provides a quantum process tomography method based on low-rank matrix sensing.

\end{remark}


\subsection{Blockwise measurement design}\label{sec_one_row}
In the previous setup, different observables were used for each initial state. In our second measurement design, we assume that the same class of observables is applied across all initial states. With an appropriate selection of initial states and observables, along with a mild assumption on $\bK$, this approach enables efficient optimization while maintaining the optimal number of measurements.

Suppose we take $E_{lk}$ to be the matrix with one on its $(l, k)$-th entry and zero elsewhere, where $k, l\in[N]$. Assume for now that $\rho_0 = E_{lk}$ and for an arbitrary observable $O$,  the first equality in Proposition \ref{prop_measurement_tensor_product} implies 
\begin{equation}
    \innerp{\rho_1, O}_{\C^{N \times N}} = \innerp{\rho_0^\top\otimes O^\dagger, \bK^\dagger}_{\C^{N^2 \times N^2}} = \innerp{E_{kl}\otimes O, \bK}_{\C^{N^2 \times N^2}} = \innerp{O, \bK_{kl}}_{\C^{N^2 \times N^2}},
\end{equation}
where $\bK_{kl}$ is the $(k, l)$-th block of the matrix $\bK$ when divided to $N\times N$ submatrices in $\C^{N \times N}$. If we can have data for $\innerp{O, \bK_{kl}}$, we can isolate the tomography of $\bK$ onto its sub-blocks $\bK_{kl}$. However, $E_{lk}$ is not a valid density matrix for $ k\neq l$, so the data is not directly available to us. Instead, we consider four initial states,  
\begin{equation}\label{eq_4_states_Elk}
	\begin{aligned}
		\rho_0^{kl} = \frac{1}{2}(E_{kl} + E_{lk} + E_{kk} + E_{ll}),\quad 
		\rho_0'^{kl}= \frac{1}{2}(iE_{kl} -i E_{lk} + E_{kk} +E_{ll}),\quad 
		\rho_0^{kk} = E_{kk}, \quad \rho_0^{ll} = E_{ll},
	\end{aligned}
\end{equation}
so that 
\begin{equation}\label{eq_obserbable_Elk}
	E_{lk} = (\rho_0^{kl} + i\rho_0'^{kl }) - \frac{1 + i}{2}(\rho_0^{kk} + \rho_0^{ll}).
\end{equation}
We are now able to equivalently have initial state $\rho_0 = E_{lk}$ with the price of requiring a constant multiple of the original number of initial states. When $k = l$, we just let $\rho_0 = E_{kk}$ directly.

\begin{remark}
Note that the reshaping operator $\mR$ is designed such that Proposition \ref{prop_measurement_tensor_product} holds when the right-hand side takes an inner product with $\overline{\rho_0} \otimes O$. If we define the reshaping operator as 
\begin{equation}\label{eq_mR_old}
	\mR_1(A) = 
	\begin{bmatrix}
		\widetilde A_1\\ \vdots \\ \widetilde A_{n}	
	\end{bmatrix}
	,\quad 
	\widetilde A_j = 
	\begin{bmatrix}
		\vec(A_{j1}^\top)^\top \\ 
		\vdots \\
		\vec(A_{jn}^\top)^\top	
	\end{bmatrix},
\end{equation}
which has the property $\mR_1(B \otimes C) = \vec(B^\top) \vec(C^\top)^\top$, so that an equivalent result of Proposition \ref{prop_measurement_tensor_product} gives $\innerp{\rho_1, O}_{\C^{N\times N}} = \innerp{\overline{O}\otimes \rho_0, \bK}_{\C^{N^2\times N^2}}$. To achieve a similar reduction to subblocks, one only needs to introduce two observables so that their linear combination is $E_{ij}$, which is less than the number of initial states used since observables do not have the requirement of trace one. However, the density matrices $\rho_0$ must be random Pauli matrices for the theoretical guarantee of our proposed method. While Remark \ref{rmk_Pauli_eigen} offers an alternative using eigendecomposition, it comes at the cost of additional computation overhead. Nevertheless, numerical tests suggest that randomly generated density matrices and observables often outperform Paulis for the learning task. We remark on this alternative reshaping operator definition, which is essentially due to state-observable equivalence, as it may be useful in some experimental settings.
\end{remark}

Suppose we have observables $\{O^m\}_{m = 1}^{M_O}$, where $M_O$ denotes the number of observables since we have shared observables and initial states. We define the measurement operator and the result as
\begin{align}\label{eq_A_b_kl}
    \widetilde{\mA} (\bK_{kl}) = 
    \begin{bmatrix}
	\tr[{\bA}_1^\dagger\bK_{kl} ]\\
	\vdots \\ 
	\tr[{\bA}_{M_O}^\dagger\bK_{kl}  ]\\
	\end{bmatrix}, 
	\quad 
	\bb_{kl} = \bb^\star_{kl} + \bxi_{kl}
	\in \C^{M_O}
\end{align}
with ${\bA}_m = O^m$, $\bb^\star_{kl} = [b^{\star 1}_{kl}, \dots, b^{\star {M_O}}_{kl}]^\top$ where 
\begin{equation}
	\quad b^{\star m}_{kl} = \tr[\mK^\star(E_{lk})^\dagger O^m] = \tr\left[ {O^m } \bK_{kl}^\star\right], 
\end{equation}
and $\bxi_{kl} = [\xi^1_{kl}, \dots, \xi^{M_O}_{kl}]^\top$ where $ \xi^{m}_{kl} \overset{\mathrm{iid}}{\sim} \mN(0, \sigma^2)$. Note that $\mK^\star(E_{lk})$ used the linear combination of $\mK^\star$ on the four initial states in \eqref{eq_4_states_Elk}. We could define the following matrix sensing objective
\begin{equation}\label{eq_loss_kl}
	   \quad \argmin_{U_{kl}, V_{kl} \in \C^{N \times r},\ \bK_{kl} = U_{kl} V_{kl}^\dagger}f_{kl}(\bK_{kl}), \quad f_{kl}(\bK_{kl}) = \frac{1}{2{M_O}}\norm{\widetilde{\mA}(\bK_{kl}) - \bb_{kl}}^2.\
    \tag{$\mQ_{kl}$}
\end{equation}
If we assume that $r \leq N$,
this problem remains a low-rank matrix sensing task since $\rank(\bK_{kl}) \leq \rank(\bK) = r$. However, it is possible that $r \geq N$ and we postpone the discussion of $r \geq N$ to Section \ref{sec_K_blocks_conclusion}.

The sensing operator $\widetilde{\mA}$ constructed from $\{O^m\}_{m=1}^{M_O}$, resembles quantum state tomography, though the target matrix $\bK_{kl}$ is generally neither Hermitian nor a trace-one quantum state. Since the same operator $\widetilde\mA$ is used for all $(k, l)$ pairs, if $\widetilde\mA$ satisfies the RIP condition, achievable by choosing random scaled Pauli matrices $O^m$ (Theorem \ref{thm_Pauli_RIP}), the regularized form of \eqref{eq_loss_kl} has no spurious local minima for any $k$ and $l$. This measurement design decomposes the original problem \eqref{eq_loss_all} into smaller subproblems, enabling parallel recovery of blocks for improved efficiency. While full tomography involves considering all index pairs, it exceeds the required measurements due to the low-rank nature of $\bK$. Instead, it suffices to solve a subset of subproblems. Specifically, we first recover the blocks in the first row, $\bK_1 = [\bK_{11}, \dots, \bK_{1N}]$, and then reconstruct the full matrix $\bK$ from these blocks. In fact, any row can be selected, provided the chosen block rows have rank $r$ (see Assumption \ref{asmp_block_rank}).



We will first present the algorithms for learning the blocks on the first row in Section \ref{sec_K_first_row} with theoretical performance guarantee using the matrix sensing framework. Then, we introduce the assumptions needed to reconstruct the matrix $\bK$ in Section \ref{sec_K_assumption}. We introduce the detailed reconstruction of $\bK$ from first-row blocks in Section \ref{sec_K_rest_from_first_row} and conclude in Section \ref{sec_K_blocks_conclusion}.



\subsection{Learning the first-row blocks}\label{sec_K_first_row}
We start with learning the blocks in the first row, which will need to equivalently evaluate the initial states as $E_{11}, \dots, E_{1N}$. From the design of the initial states in \eqref{eq_4_states_Elk}, we need $3N-2$ initial states in total. To obtain data $\bb_{11}, \dots ,\bb_{1N}$, we need $M_O$ observables and therefore $M = M_O(3N-2)$ measurements. 


\subsubsection{Parallel ALS}\label{sec_Parallel_ALS}
\begin{algorithm}
\caption{Parallel ALS.}\label{alg_parallel_ALS}
    \hspace*{\algorithmicindent} 
    \textbf{Input:} Observables $\{O^m\}_{m = 1}^M$, observation data $\bb_{11}, \dots, \bb_{1N}$, rank $r$.\\    \hspace*{\algorithmicindent} 
    \textbf{Output:} Estimated blocks $\bK_{11}, \bK_{12}, \dots \bK_{1N}$.\\
\begin{algorithmic}[1]
\FOR{$k = 1, \dots, N$}
	\STATE{Estimate $\bK_{1k}$ using Nesterov-ALS (Algorithm \ref{alg_ALS_N}) with sensing operator $\widetilde{\mA}$ and measurements $\bb_{1k}$ defined in \eqref{eq_A_b_kl}}. 
\ENDFOR
\RETURN{$\bK_{11}, \dots, \bK_{1N}$.}
\end{algorithmic}
\end{algorithm}
We first use ALS in parallel to learn each block $\bK_{1k}$ for $k \in [N]$. The algorithm is summarized in Algorithm \ref{alg_parallel_ALS}. Note that the algorithm's performance depends on the estimation of each block $\bK_{1k}$, which is a low-rank matrix sensing problem with sensing operator $\widetilde{\mA}$, observations $\bb_{1k}$ and dimension $N$. Taking the observables to be iid samples of random scaled Pauli matrices, we have the performance guarantee from Theorem \ref{thm_incoherence_RIP_landescape} with $d = N$.

\begin{corollary}
	Suppose $\widetilde{\mA}$ and $\bb_{1k}$ are defined in \eqref{eq_A_b_kl}, where $\{O^m\}$ are iid samples of scaled Pauli matrices. Then for any $k \in [N]$, if $M_O \geq C r N \log^6 N$ where $C= O(1/\delta^2)$ is a constant, $\frac{1}{c}\widetilde\mA$ satisfies the $(r, \delta)$-RIP condition with high probability over the selection of the matrices, with $c$ and $\delta$ defined in \eqref{eq_delta_c_RIP}. Moreover, if $\delta \leq \frac{1}{20}$, the following hold for the matrix sensing objective \eqref{eq_loss_ALS_general_reg} using $\widetilde{\mA}$ and $\bb_{1k}$ with $\lambda = c/8$.
	\begin{enumerate}
		\item For the noiseless case $(\sigma = 0)$ , all local minima $(U, V)$ satisfies $UV^\dagger  = \bK_{1k}^\star$. 
		\item For the noisy case $(\sigma > 0)$, all local minima must satisfy 
	\[
	    \| U V^\dagger - \bK_{1k}^\star \|_{\C^{N\times N}} \leq O\left(\sigma N^2\sqrt{\frac{rN\log M_O}{M_O}}\right)
	\]	
	with high probability.
	\end{enumerate}
\end{corollary}
%
%
%
%
%
%
%

Although parallel learning offers computational efficiency, we comment that this algorithm does not exploit the low-rank structure of the first row since each block is learned independently and lacks a shared column space.

\subsubsection{Joint ALS for the first row}\label{sec_ALS_N}
Next, we present an algorithm that combines the learning of all blocks in the first row. Consider the stacked sensing operator $\widetilde\mA^{N} = \oplus_{k = 1}^N \widetilde{\mA}$ and measurements $\bb_1$ where
\begin{equation}\label{eq_A_b_1}
	\widetilde\mA^{N}(\bK_{1}) = 
	\begin{bmatrix}
 		\widetilde{\mA}(\bK_{11})\\
 		\vdots\\
		\widetilde{\mA}(\bK_{1N})
	\end{bmatrix}
	,\quad
	\quad
	\bb_1 = 
	\begin{bmatrix}
 		\bb_{11}\\
 		\vdots\\
		\bb_{1N}
	\end{bmatrix}
    \in \C^{NM_O},
\end{equation}
and the matrix sensing objective 
\begin{equation}\label{eq_loss_1}
	   \quad \argmin_{U\in \C^{N \times r},\ V \in \C^{N^2 \times r}, \ \bK_1 = UV^\dagger}\frac{1}{2NM_O}\norm{\widetilde{\mA}^N(\bK_{1}) - \bb_{1}}^2.\
    \tag{$\mQ_{1}$}
\end{equation}
Note that this is different from a simple stacking of the subproblems, as the same column space of $\bK_{1k}$ is shared by using the same parameter $U$. We apply Nesterov-ALS (Algorithm \ref{alg_ALS_N}) with $\widetilde{\mA}^N$ as the sensing operator and $\bb_1$ as the measurements. The performance of the algorithm is characterized by the RIP condition of $\widetilde{\mA}^N$, which is derived from the RIP condition of $\widetilde{\mA}$. 

\begin{lemma}\label{lem_AN_RIP}
	Suppose $\widetilde{\mA}$ satisfy the $(r, \delta)$-RIP condition, so does $\widetilde{\mA}^N$.
\end{lemma}
\begin{proof}
	For any matrix $\bX \in \C^{N \times N^2}$ with rank $r$, we first decompose $\bX$ into $N \times N$ blocks with rank as most $r$, $\bX = [\bX_1, \dots, \bX_N]$. 
	By the RIP condition of $\widetilde \mA$, $(1-\delta)\norm{\bX_k}_{C^{N \times N}}^2 \leq \norm{\widetilde\mA \bX_k}^2 \leq (1+ \delta) \norm{\bX_k}_{\C^{N \times N}}^2$ for any $k = 1,\dots, N$. Then 
	\begin{equation}
		\norm{\widetilde{\mA}^N(\bX)}^2 = \sum_{k = 1}^N\norm{\widetilde\mA \bX_k}^2 \leq (1+\delta) \sum_{k = 1}^N \norm{\bX_k}_{\C^{N \times N}}^2 = (1+\delta)\norm{\bX}_{\C^{N \times N^2}}^2 .
	\end{equation}
	and the other side of the inequality follows similarly. 
\end{proof}

\begin{remark}\label{rmk_stack_RIP}
	Since a slightly different version of the RIP was originally used in \cite{liuUniversalLowrankMatrix2011}, we need to prove that $\widetilde\mA^N$ satisfy \eqref{eq_RIP_Pauli_different} when $\widetilde\mA$ does. Nevertheless, the same reasoning holds as in Lemma \ref{lem_AN_RIP}.
\end{remark}

Again, the RIP condition of $\widetilde\mA$ is achieved if we take iid random scaled Pauli observables, and we can characterize the optimization landscape as follows. 

\begin{corollary}\label{cor_Q1_guarantee}
	Suppose $\widetilde{\mA}^N$ and $\bb_{1}$ are defined in \eqref{eq_A_b_1}, where $\{O^m\}$ are iid samples of scaled Pauli matrices. Then for $M_O \geq C r N \log^6 N$ where $C= O(1/\delta^2)$ is a constant, $\frac{1}{c}\widetilde{\mA}^N$ satisfies the $(r, \delta)$-RIP condition with high probability over the selection of the matrices, with $c$ and $\delta$ defined in \eqref{eq_delta_c_RIP}. Moreover, if $\delta \leq \frac{1}{20}$, the following hold for the matrix sensing objective \eqref{eq_loss_ALS_general_reg} using $\widetilde{\mA}^N$ and $\bb_{1}$ with $\lambda = c/8$. 
	\begin{enumerate}
		\item For the noiseless case $(\sigma = 0)$, all local minima $(U, V)$ satisfies $UV^\dagger  = \bK_{1}^\star$. 
		\item For the noisy case $(\sigma > 0)$, all local minima must satisfy 
	\[
	    \| U V^\dagger - \bK_{1}^\star \|_{\C^{N\times N^2}} \leq O\left(\sigma N^2\sqrt{\frac{rN\log (NM_O)}{M_O}}\right)
	\]	
	with high probability.
	\end{enumerate}

\begin{proof}
By the Remark \ref{rmk_stack_RIP} and Lemma \ref{lem_AN_RIP}, $\frac{1}{c}\widetilde \mA^N$ satisfies $(r, \delta)$-RIP with $c = \frac{1 + \widehat{\delta}^2}{N^2}$. Then by Theorem \ref{thm_landscape_noisy} with $d_1 = N$, $d_2 = N^2$ and $M = NM_O$, and $\sigma = \frac{\sigma}{c}$, we have the desired result. \end{proof}


%
%
%
%
\end{corollary}


In this approach, the column space is shared. However, since the matrix $V\in \C^{N^2\times r}$, the least square for solving $V$ would be expensive. That suggests we select a subset of the blocks on the first row to be learned jointly.

\subsubsection{Joint ALS for the subset of the first row}\label{sec_ALS_I}
We select an index set $I = [i_1, \dots ,i_{p}]$ of total length $p$, where $i_1 = 1$ and $i_2, \dots i_{p}$ are random samples from $2, \dots, N$ without replacement. We fix the position $i_1 = 1$ since the first block $\bK_{11}$ is Hermitian. Stack the blocks on the first row corresponding to $I$, where $\bK_I = [\bK_{1i_1},\dots,\bK_{1i_p}] \in \C^{N \times Np}$. We define the stacked sensing operator $\widetilde\mA^I = \oplus_{k = 1}^p \widetilde \mA$ and measurements $\bb_I$, 
\begin{equation}\label{eq_A_b_I}
	\widetilde\mA^I(\bK_I) = \begin{bmatrix}
		\widetilde\mA(\bK_{1i_1})\\
		\vdots\\
		\widetilde\mA(\bK_{1i_p})
	\end{bmatrix},\quad 
	\bb_I = \begin{bmatrix}
		\bb_{1i_1}\\
		\vdots \\
		\bb_{1i_p}
	\end{bmatrix}
	 \in \C^{M_O p}.
\end{equation}
We consider the matrix sensing problem 
\begin{equation}\label{eq_loss_I}
	   \quad \argmin_{U\in \C^{N \times r},\ V \in \C^{Np \times r}, \ \bK_I = UV^\dagger}  \frac{1}{2M_O p}\norm{\widetilde{\mA}^I(\bK_{I}) - \bb_{I}}^2,\
    \tag{$\mQ_{I}$}
\end{equation}
which can be solved using ALS-Nesterov (Algorithm \ref{alg_ALS_N}). Recall the problem \eqref{eq_loss_kl}, when $k = 1$, the decomposition $U$ is shared for $l = 1, \dots, N$. Therefore, using the matrix $U$ derived from \eqref{eq_loss_I}, we can solve for $V_{1l}$ for the rest indices $l \notin I$ using least squares directly. The algorithm is summarized in Algorithm \ref{alg_ALS_I}.

\begin{algorithm}
\caption{Joint ALS for a subset $I$ of the first-row.}\label{alg_ALS_I}
    \hspace*{\algorithmicindent} 
    \textbf{Input:} Observables $\{O^m\}_{m = 1}^M$, observation data $\bb_{11}, \dots, \bb_{1N}$, rank $r$.\\    \hspace*{\algorithmicindent} 
    \textbf{Output:} Estimated blocks $\bK_{11}, \bK_{12}, \dots \bK_{1N}$.\\
\begin{algorithmic}[1]
\STATE{Estimate $\bK_I = UV^\dagger$  by solving \eqref{eq_loss_I} using ALS-Nesterov (Algorithm \ref{alg_ALS_N}) with sensing operator $\widetilde\mA^I$ and measurements $\bb_I$. Denote $V^\dagger = [V_{1i_1}^\dagger, \dots, V_{1i_p}^\dagger]$.}
\FOR{$k = 1, \dots, N, k \notin I$}
	\STATE{Estimate $V_{1k}$ by solving $\displaystyle\argmin_{V_{1k} \in \C^{N \times r}} f_{1k}(UV_{1k}^\dagger)$ defined as in \eqref{eq_loss_kl}, using least squares, with $U$ fixed to the previous estimation.} 
\ENDFOR
\RETURN{$\bK_{1k} = UV_{1k}^\dagger, k = 1,\dots, N$.}
\end{algorithmic}
\end{algorithm}

\begin{remark}\label{rmk_ALS-I}
    The performance of a regularized version of the matrix sensing problem \eqref{eq_loss_I} relies on the RIP condition of $\widetilde\mA^I$. Since the analysis presented in Lemma \ref{lem_AN_RIP} and Remark \ref{rmk_stack_RIP} applies directly to $\widetilde\mA^I$, the same performance guarantees hold for problem \eqref{eq_loss_I}. However, achieving a comparable performance guarantee for the recovery of  $\bK_1$ requires estimating the error in $U$. To address this, we can use the estimated $U$ as an initial value and reapply ALS to problem \eqref{eq_loss_1}, ensuring the same result as stated in Corollary \ref{cor_Q1_guarantee}. In practice, we directly implement Algorithm \ref{alg_ALS_I}, as empirical results demonstrate its effectiveness in achieving accurate recovery.
\end{remark}


Given the sufficient number of observables for the first-row blocks $M_O \geq CrN \log^6 N$, we need the number of total measurements $M \geq CrN^2\log^6N$ for a successful recovery of the superoperator, which agrees with the order in the random measurement design.
In Section \ref{sec_Numerics}, we numerically justify the necessary number of measurements and compare the performance and complexity of these three methods, together with ALS in random measurement design. Empirically, Algorithm \ref{alg_ALS_I} outperforms the others in a joint consideration of accuracy and computation time.

\subsection{Blockwise matrix completion}\label{sec_K_assumption}
%
%

The second step involves reconstructing the entire matrix $\bK$ from the estimated submatrices $\bK_{kl}$. This process resembles the matrix completion problem but observes matrix blocks instead of individual entries. It allows us to impose a similar structural requirement as in matrix completion.
\begin{assumption}\label{asmp_block_rank}
    Let $\bK^\star$ be the reshaped matrix of a quantum operator $\mK^\star$. We assume that the first block matrix $\bK^\star_{11}$ has rank $r$.
\end{assumption}
This assumption can fail, for instance, with a diagonal low-rank matrix where off-diagonal blocks are zero, making reconstruction impossible without observing the $r$ diagonal elements. Nonetheless, it parallels the incoherence condition in matrix completion \cite{candes2012exact}, which prevents the singular vectors of the underlying matrix from being too aligned with the standard basis vectors.

\begin{remark}
In matrix completion, the goal is to recover a low-rank matrix $\bX$ from a subset of its entries. 
The incoherence condition ensures that the singular vectors of $\bX$ are spread out across all coordinates rather than being concentrated in a few. 
This 
guarantees that each entry of the matrix contains a balanced amount of information, enabling successful reconstruction from a randomly selected subset of entries. Note that this is a prior condition on $\bX$, rather than the requirement for sensing matrices as in Definition \ref{def_incoherent_matrix}.


\end{remark}



We shall prove that Assumption \ref{asmp_block_rank} holds with probability one for low-rank matrix $\bK$ constructed from randomly chosen unitary matrices following Haar measure \cite{diaconis1994eigenvalues}. The Haar measure is a translation-invariant measure defined on locally compact groups such as the unitary group $\mathcal{U}(N)$. Its uniqueness makes it a standard notion of uniform distribution among these groups. Specifically, for the unitary group, the Haar measure allows for the definition of uniformly random unitary matrices. 


\begin{lemma}\label{lem_Haar_block}
Suppose $\bU$ is a uniformly random unitary matrix in $\C^{N^2\times N^2}$ with respect to the Haar measure. Let $\bU_{kl}$ denote the $(k, l)$-th subblock of $\bU$. Then 
for any $k, l = 1, \dots, N$, $\bU_{kl}$ is invertible with probability 1. 
\end{lemma}

\begin{proof}
Let $\mS_{kl} = \left\{
\bU\in\C^{N^2\times N^2} \text{ is unitary} \mid \rank(\bU_{kl}) < N  \right\}$. We aim to show that the Haar measure of $\mS_{kl}$ is zero. Consider the function $F_{kl}(\bU) = \det(\bU_{kl})$. The set $\mS_{kl}$ corresponds to the preimage of zero under $F_{kl}$, i.e., $\mS_{kl} = F_{kl}^{-1}(0)$. Since $F_{kl}(\bU)$ is a polynomial in the entries of $\bU$, the set of its roots forms an algebraic variety, which has Lebesgue measure zero unless the polynomial is identically zero.

However, $F_{kl}(\bU)$ is not identically zero because there exist unitary matrices for which $\bU_{kl}$ is invertible. Since the Haar measure is absolutely continuous with respect to the Lebesgue measure on the unitary group, the set $\mS_{kl}$ has Haar measure zero. Therefore, $\bU_{kl}$ is invertible with probability 1.
\end{proof}

%
%
%
%
%

Suppose the true operator matrix has eigendecomposition \( \bK^\star = \bP^\star \bD \bP^{\star \dagger} \), where $\bD^\star$ is an $ r \times r$ diagonal matrix and $\bP^\star= [\bP_{1}^\star, \dots, \bP_{N}^\star] \in \C^{N^2 \times r}$ is divided into $N \times r$ blocks. Assume that \( \bP^\star \) consists of the first \( r \) columns of a unitary matrix \( \bU \) sampled according to the Haar measure. By Lemma \ref{lem_Haar_block}, the blocks of \( \bP^\star \) have rank \( r \) with probability 1. Since $\bK_{kl} = \bP^{\star}_k \bD^\star \bP_l^{\star\dagger}$, Assumption \ref{asmp_block_rank} is satisfied almost surely under Haar probability measure. 

\begin{remark}\label{rmk_1_rank_r}
	Lemma \ref{lem_Haar_block} actually provides a much stronger result than Assumption \ref{asmp_block_rank}. In particular, any subblock $\bK_{kl}$ of $\bK$ has rank $r$. Moreover, we shall see that we only need there exists at least one $k \in [N]$ so that  $\rank(\bK_{kk})  = r$. Then, we can transfer the previous algorithms onto the $k$-th row and then recover the entire matrix $\bK$ using the following steps. 
\end{remark}



\subsection{Deterministic reconstruction from the first row}\label{sec_K_rest_from_first_row}
We are left to reconstruct the entire matrix $\bK$ from its first row $\bK_1$. Note that $\bK$ is Hermitian. From the blocks on the first row, we can derive the blocks on the first column, $\bK_{k1} = \bK_{1k}^\dagger$. Then we apply randomized SVD to $\bK_1$ \cite{martinsson2011randomized}, which is a fast and efficient algorithm that approximates the singular value decomposition of large matrices by leveraging random projections to reduce dimensionality before applying standard SVD techniques. Denote the result as 
\begin{equation}\label{eq_rsvd_K1}
	\bU \bS \bJ^\dagger = \bK_1, \quad \bU \in \C^{N \times r}, \bS \in \C^{r \times r}, \bJ \in \C^{N^2 \times r},
\end{equation}
where
\begin{equation}
	\bJ = 
	\begin{bmatrix}
 		\bJ_1\\
 		\vdots\\
 		\bJ_N
 	\end{bmatrix},
	\quad \bJ_k \in \C^{N\times r}, \quad k =1 ,\dots, N.
\end{equation}
By the Assumption \ref{asmp_block_rank}, $\bK_1$ has rank $r$. Therefore, $\bJ^\dagger$ consists of the row basis vectors of $\bK$. Again, using Assumption \ref{asmp_block_rank}, the first block $\bJ_1$ has rank $r$, we can find the coefficients of the block $\bK_{k1}$ over the row block $\bJ_1^\dagger$, 
\begin{equation}\label{eq_Ck}
	\bC_k = \bK_{k1}  (\bJ_{1}^\dagger)^{-1},
\end{equation}
where $(\bJ_1^\dagger)^{-1} \in \C^{N \times r}$ represents the pseudo inverse of $\bJ_{1}^\dagger$. At last, we use the coefficient matrix $\bC_k \in \C^{N \times r}$ to reconstruct the $k$-th block row of $\bK$,
\begin{equation}\label{eq_Kk}
	\bK_k = \bC_k \bJ^\dagger,
\end{equation}
where $\bK_k = [\bK_{k1}, \dots, \bK_{kN}] \in \C^{N \times N^2}$. Note that the derived estimator $\bK$ has rank $r$ since all its rows are linear combinations of row basis vectors $\bJ$. The procedure is summarized in Algorithm \ref{alg_det_reconstruction}.


\begin{algorithm}
\caption{Deterministic reconstruction from blocks.}\label{alg_det_reconstruction}
    \hspace*{\algorithmicindent} \textbf{Input:} Estimated blocks $\bK_{11}, \bK_{12}, \dots \bK_{1N}$, rank $r$.\\    \hspace*{\algorithmicindent} \textbf{Output:} Estimated reshaped operator matrix $\bK$.\\
\begin{algorithmic}[1]
\STATE{Find the first $r$ row basis vectors $\bJ$ of $\bK_1$ using randomized SVD as in \eqref{eq_rsvd_K1}.}
\STATE{Compute the pseudo inverse $(\bJ_1^\dagger)^{-1}$.}
\FOR{$k = 2, \dots, N$}
	\STATE{Assign $\bK_{k1} = \bK_{1k}$.}
    \STATE{Obtain the coefficients $\bC_{k} \in \C^{N \times r}$ of $\bK_{k1}$ over row vectors $\bJ_{1^\dagger}$ using \eqref{eq_Ck}.}
    \STATE{Reconstruct the $k$-th row block $\bK_k$ using \eqref{eq_Kk}.}
\ENDFOR
\RETURN{$\bK$.}
\end{algorithmic}
\end{algorithm}

\begin{remark}[Performance guarantee]\label{rmk_error_det_reconstruction}
	The error of the estimated matrix $\bK$ depends on the error in $\bK_{1k}$, the condition number of $\bJ_1$, and the error in $\bJ$. The error in $\bJ$ further depends on the spectral gap and the error in $\bK_1$ (Davis–Kahan Theorem, \cite{davis1970rotation}). In practice, it is difficult to determine the condition number of $\bJ_1$ and the spectral gap of $\bK_1$ apriori. However, as noted similarly in Remark \ref{rmk_ALS-I}, $\bJ$ can be an initial value for a joint ALS over $\bK$. While the measurement setting may not ensure a good RIP condition for $\mA_0$, introducing additional random measurements can establish RIP, enabling the landscape characterization in Corollary \ref{cor_ALS_N_2}. Numerical results show that the deterministic reconstruction Algorithm \ref{alg_det_reconstruction} performs well in practice.


\end{remark}

\begin{remark}[Complexity analysis]
	The complexity of the randomized SVD is $O(mn\log(r) + r^2(m+n))$, where $m$ and $n$ denote the dimensions of the matrix, and $r$ represents its rank \cite{halko2011finding}. In our case, \(\bK_{1} \in \C^{N \times N^2}\), resulting in a complexity of $ O(N^3\log(r) + N^2r^2)$ compared to the direct SVD with a complexity of \( O(N^4) \). Computing the pseudo-inverse \((\bJ_{1}^\dagger)\) has a complexity of \( O(Nr^2) \), which follows directly from the complexity of SVD. The matrix multiplications in \eqref{eq_Kk} require \( O(N^3r) \), which dominates the overall complexity of computing \(\bK_k\). Consequently, the total complexity of the algorithm is \( O(N^4r) \), as there are \( N-1 \) block rows, and the iterations over rows can be executed in parallel.
\end{remark}

\begin{remark}
	One could have a similar reconstruction of $\bK$ using the SVD for only the block $\bK_{11}$. Find the project coefficient $\bC_{k1}$ of the rows of $\bK_{k1}$ onto the row basis vectors $\bK_{11}$, and then reconstruct $\bK_{kl}$ using the coefficient matrix $\bC_{k1}$ on the transformed rows of $\bK_{1l}$, where the transformation expresses the rows of $\bK_{11}$ into its row basis vectors. This method will have a better parallel potential, but it fails to preserve the low-rank structure of $\bK$ since it does not apply the same basis row vectors in the reconstruction. See also the discussion in Section \ref{sec_Parallel_ALS}.
\end{remark}

\subsection{Summary and extensions}\label{sec_K_blocks_conclusion}
We summarize our approach here: First, we estimate the first row blocks using parallel ALS, joint ALS, or joint ALS for a subset of the first row. Then we apply Algorithm \ref{alg_det_reconstruction} to recover the entire matrix $\bK$. 
We also discuss some extensions below. 


\subsubsection{Group Synchronization}
Selecting the first row is an example and not a requirement, as shown in Algorithm \ref{alg_det_reconstruction}. If it is not possible to choose the initial states as in \eqref{eq_4_states_Elk} to measure $\bK$ in any of its rows, the measurement design can be relaxed by randomly selecting $E_{kl}$. The group synchronization technique can then reconstruct $\bK$ from the randomly selected $\bK_{kl}$.

	Since $\bK^\star$ is Hermitian, we can find decomposition so that $\bK^\star = \bW^\star \bE \bW^{\star\dagger}$, where $\bE \in \C^{r \times r}$ is a diagonal matrix with $r_+$ ones and $r_-$ negative ones and $\bW^\star \in \C^{N^2 \times r}$. We decompose $\bW^{\star\dagger} = [\bW^{\star\dagger}_1, \dots, \bW^{\star\dagger}_N]$ where $\bW^\star_k \in \C^{N \times r}$. Note that for any indices $k, l \in [N]$, $\bK_{kl} = \bW^\star_{k} \bE \bW_{l}^{^\star\dagger}$.
	
	First, we construct the density matrices $E_{11}, \dots, E_{NN}$ in the measurement setting \eqref{eq_A_b_kl} to estimate $\bK_{11}, \dots, \bK_{NN}$. Since these matrices are Hermitian, we can express them in the form $\bK_{kk} = \bW_k \bE \bW_k^\dagger$. However, $\bW_k$ may differ from $\bW_k^\star$ by a group element. Specifically, there exists some $\bG_k \in \C^{r \times r}$ such that $\bW_k^\star = \bW_k \bG_k$ and $\bG_k \bE \bG_k^\dagger = \bE$. The matrices $\bG_k$ consist the pseudo-unitary group $U(r_+, r_-)$ (See \cite{munshi2019self} and references therein). The task then reduces to determining $\bG_1, \dots, \bG_N$ to reconstruct $\bK$. Clearly, information from the diagonal blocks alone is insufficient for this reconstruction.

	Suppose we have an off-diagonal estimation of $\bK_{kl}$, by the result on the diagonal blocks, we have 
	\begin{equation}
		\bW_k \bG_k \bE \bG_l^\dagger \bW_l^\dagger = \bK_{kl}.
	\end{equation} 
	By Lemma \ref{lem_Haar_block}, we can assume that $\rank(\bW_l) = \rank(\bW_k) = r$. Therefore we can apply the pseudo inverse of $\bW_k$ and $\bW_l$ from both sides, so that
\begin{equation}
	\bG_k \bE \bG_l^\dagger = \bW_k^{-1} \bK_{kl} \bW_{l}^{-\dagger} := \bC_{kl}
\end{equation}
The matrix $\bC_{kl}$ is called a quadratic measurement of the group elements $\bG_1, \dots \bG_N$. In the case that $\bE$ is the identity matrix, this becomes a classical synchronization problem in a unitary group $U(r)$ \cite{wang2013exact}. For the general pseudo unitary group where $r_- \geq 1$, we can apply the approach in \cite{liu2023unified} to transform this problem into matrix completions. 

\subsubsection{The case of $r \geq N$}\label{sec_3_r_geq_N}
The estimation of the subblock $\bK_{kl}$ remains a low-rank sensing problem that requires that $r \leq N$, and we can extend the method similarly to the case of $r \geq N$. For example, assume that $N \leq r \leq 2N$, we could merge the blocks $\bK_{1:2,1:2} = [\bK_{11}, \bK_{12}; \bK_{21}, \bK_{22}]$ and consider a larger block system with block size $2N \times 2N$. Similar initial states and observables can be constructed to isolate the tomography of the blocks like $\bK_{1:2,1:2}$. We leave this a future research direction.

%
%
%
%
%
%
%

\section{Numerical results}\label{sec_Numerics}
In this section, we show numerical examples that validate our theory. We first introduce representative examples of learning quantum channels and Lindbladians. Second, we show the recovery rate and computation time scaled with the number of data. We also show the error decay with data size in the noisy case. At last, we give the necessary data size scaling with the dimension of Hilbert space $N$ and rank $r$. 


In the following examples, we generate data using Qutip in Python \cite{johansson2012qutip}. The initial states are generated using \texttt{rand\_dm}, which provides random density matrices. We use \texttt{rand\_herm} to generate random Hermitian observables and Hamiltonian for the Lindbladian. The jump operators are randomly generated and have iid Gaussian entries. The quantum channels are generated using \texttt{rand\_super\_bcsz}, which applies the method in \cite{bruzda2009random} to ensure it is CPTP. Random scaled Pauli matrices are iid samples of the form in Example \ref{ex_scaled_Pauli}.

The relative Frobenius error is defined as 
\begin{equation}
	\frac{\norm{\bK - \bK^\star}_{\C^{N^2\times N^2}}}{\norm{\bK^\star}_{\C^{N^2\times N^2}}}.
\end{equation}
The diamond norm is defined as 
\begin{equation}
	\norm{\mK}_\diamond := \norm{\mK \otimes \id_n}_{1 \to 1} = \sup_{A} \frac{\norm{\mK(A)}_1}{\norm{A}_1},
\end{equation}
where $A$ is selected among $N\times N$ positive semidefinite matrices and $\norm{A}_1:= \tr(\sqrt{A A^\dagger})$ is the trace norm.  The diamond norm is a meaningful and widely used metric for quantum processes \cite{gilchrist2005distance}. It can be computed efficiently using semidefinite programming \cite{watrous2012simpler, ben2009complexity, watrous2009semidefinite}. We used a package in Matlab \cite{matlab_diamond_norm}, which applied the method in \cite{watrous2012simpler} to compute the diamond norm. However, since the computation of the diamond norm is expensive, we provide the diamond norm in Section \ref{sec_representative_example} and compare the relative Frobenius norm for the rest of the examples. The numerical experiments are carried out in Matlab 2024a on an M3 Pro chip with 18 GB of memory and  11 cores. 

We compare the performance of the following methods. First, assuming a random measurement design in Section \ref{sec_3_random_measurement}, we have
\begin{itemize}
	\item ALS-$N^2$: ALS (Algorithm \ref{alg_ALS_N}) for $\mA$ and $\bb$.
\end{itemize}
Then, using the blockwise measurement design, we can learn the blocks on the first row and then recover the rest of the blocks using deterministic reconstruction (Algorithm \ref{alg_det_reconstruction}). We distinguish the following three methods for learning the blocks on the first row, 
\begin{itemize}
	\item ALS-P: The parallel learning (Algorithm \ref{alg_parallel_ALS}) of each block on the first row (Section \ref{sec_Parallel_ALS}). 
	\item ALS-N: Joint ALS for the first row (Algorithm \ref{alg_ALS_N}) with $\widetilde\mA^N$ and $\bb_1$ (Section \ref{sec_ALS_N}). 
	\item ALS-\,I\,: Joint ALS for a subset of the first row (Algorithm \ref{alg_ALS_I}) with $\widetilde\mA ^I$ and $\bb_I$ (Section \ref{sec_ALS_I}). 
\end{itemize}

Recall that $M_O$ represents the number of observables, hence the total number of measurements $M = (3N-2)M_O$.

\begin{table}[t!]
\begin{tabular}{|cc|c|c|c|c|}
\hline
\multicolumn{2}{|c|}{Method}                                     & \multicolumn{1}{c|}{ALS-$N^2$} & \multicolumn{1}{c|}{ALS-P} & \multicolumn{1}{c|}{ALS-N}    & \multicolumn{1}{c|}{ALS-I}  \\ \hline
\multicolumn{1}{|c|}{\multirow{3}{*}{\makecell{$N = 8$ \\ $M_O = 50$}}}   & Frobenius & \textbf{5.86e-4 $\pm$2.41e-5}  & 1.30e-3 $\pm$1.52e-4       & 9.32e-4 $\pm$1.48e-4          & 1.04e-3 $\pm$1.91e-4        \\ \cline{2-6} 
\multicolumn{1}{|c|}{}                               & Diamond   & \textbf{7.26e-4 $\pm$5.16e-5}  & 1.69e-3 $\pm$2.74e-4       & 1.25e-3 $\pm$2.37e-4          & 1.38e-3 $\pm$2.88e-4        \\ \cline{2-6} 
\multicolumn{1}{|c|}{}                               & Time (s)  & 5.680 $\pm$1.346               & 0.337 $\pm$0.075           & 0.082 $\pm$0.010              & \textbf{0.055 $\pm$0.006}   \\ \hline
\multicolumn{1}{|c|}{\multirow{2}{*}{\makecell{$N = 16$ \\ $M_O = 120$}}}   & Frobenius & 6.86e-4 $\pm$1.40e-5           & 9.45e-4 $\pm$2.36e-5       & \textbf{6.55e-4 $\pm$2.01e-5} & 6.82e-4 $\pm$2.39e-5        \\ \cline{2-6} 
\multicolumn{1}{|c|}{}                               & Time (s)  & 312.511 $\pm$66.833            & 2.964 $\pm$0.867          & 0.670 $\pm$0.066             & \textbf{0.363 $\pm$0.015}  \\ \hline
\multicolumn{1}{|c|}{\multirow{2}{*}{\makecell{$N = 64$ \\ $M_O = 450$}}} & Frobenius & *                              & *                          & \textbf{6.50e-4 $\pm$1.19e-5} & 7.01e-4 $\pm$1.56e-5        \\ \cline{2-6} 
\multicolumn{1}{|c|}{}                               & Time (s)  & *                              & *                          & 196.327 $\pm$30.185           & \textbf{26.887 $\pm$3.864} \\ \hline
\end{tabular}
\caption{Benchmark examples for channel learning}
\label{table_channel}
\end{table}

\begin{table}[t!]
\begin{tabular}{|cc|c|c|c|c|}
\hline
\multicolumn{2}{|c|}{Method}                                     & ALS-$N^2$                     & ALS-P                & ALS-N                         & ALS-I                       \\ \hline
\multicolumn{1}{|c|}{\multirow{3}{*}{\makecell{$N = 8$ \\ $M_O = 50$}}}   & Frobenius & \textbf{7.83e-4 $\pm$6.59e-5} & 1.91e-3 $\pm$1.94e-4 & 1.44e-3 $\pm$1.86e-4          & 1.69e-3 $\pm$2.96e-4        \\ \cline{2-6} 
\multicolumn{1}{|c|}{}                               & Diamond   & \textbf{7.23e-4 $\pm$7.41e-5} & 1.96e-3 $\pm$4.57e-4 & 1.32e-3 $\pm$3.34e-4          & 1.59e-3 $\pm$3.23e-4        \\ \cline{2-6} 
\multicolumn{1}{|c|}{}                               & Time (s)  & 4.817 $\pm$0.915              & 0.340 $\pm$0.158     & 0.086 $\pm$0.012              & \textbf{0.056 $\pm$0.012}   \\ \hline
\multicolumn{1}{|c|}{\multirow{2}{*}{\makecell{$N = 16$ \\ $M_O = 120$}}}   & Frobenius & \textbf{5.77e-4} $\pm$3.03e-5          & 8.54e-4 $\pm$1.08e-4 & {5.99e-4 $\pm$7.37e-5} & 6.29e-4 $\pm$7.63e-5        \\ \cline{2-6} 
\multicolumn{1}{|c|}{}                               & Time (s)  & 305.486 $\pm$22.798           & 2.152 $\pm$0.358    & 0.692 $\pm$0.040             & \textbf{0.376 $\pm$0.025}  \\ \hline
\multicolumn{1}{|c|}{\multirow{2}{*}{\makecell{$N = 64$ \\ $M_O = 450$}}} & Frobenius & *                             & *                    & \textbf{1.92e-4 $\pm$4.23e-5} & 2.11e-4 $\pm$5.72e-5        \\ \cline{2-6} 
\multicolumn{1}{|c|}{}                               & Time (s)  & *                             & *                    & 217.392 $\pm$35.446           & \textbf{30.138 $\pm$5.324} \\ \hline
\end{tabular}
\caption{Benchmark examples for Lindbladian learning}
\label{table_Lindbladian}
\end{table}

\subsection{Representative examples}\label{sec_representative_example}
In this section, we present some benchmarking examples of quantum channel learning and Lindbladian learning. 
We assume the true quantum channel has a Kraus rank of $3$ and hence $\rank(\bK^*) = 3$. For the Lindbladian, recall from Example \ref{ex_Linabladian}, by assuming $N_J = 1$ jump operator, the reshaped matrix of Lindbladian also has rank $3$. 

The initial states and observables are randomly generated when needed. The true outputs of the quantum channels and Lindbladians are computed directly using matrix multiplication. We assume the observation noise on the measurements is $\sigma_{obs} = 10^{-4}$. Note that this is a practical assumption on quantum channel tomography; however, this is not true for Lindbladian. The Lindblad equation characterizes the continuous evolution of the quantum states so that the output of the generator Lindbladian is the derivative of the trajectory. The derivative estimation is a non-trivial task. See the discussion in Section \ref{sec_related_work_Lindbladian}. Here, we focus on the low-rank structure of the Lindbladian and leave the derivative estimation as a future work. 

We list the results for three cases, $N = 8$, $N = 16$, and $N = 64$, corresponding to $n = 3, 4$ and $6$ qubits.  We repeat the test with $10$ randomly generated true superoperators and present the mean and standard deviation. We only compute the diamond norm for $N = 8$ as it is computationally expensive. For the case of $N = 64$, we only compare the performance of ALS-N and ALS-I. For ALS-I, we apply the subset ratio of $0.4$ for $N = 8$ and $N = 16$, while a ratio of $0.1$ for $N = 64$. See Table \ref{table_channel} and Table \ref{table_Lindbladian} for the results of channel learning and Lindbladian learning. In general, the results of learning channels and Lindbladians have no significant differences.

We note that ALS-$N^2$ has the best performance in terms of Frobenius error and diamond norm error because of the incoherence of the measurement inputs and a joint recovery across the blocks. ALS-P has the worst performance due to the lack of synchronization in the column space. The performance of ALS-N is slightly better than ALS-I, and both are comparable to ALS-$N^2$, especially for $N = 16$. 

A blockwise measurement design significantly improves computation time. Using subset recovery, ALS-I has the best efficiency. For $ N = 8$, it is 100 times faster than ALS-$N^2$, and it is 1000 times faster when $N = 16$. The improvement of ALS-I in efficiency compared to ALS-N depends on the ratio of choosing subsets.

\subsection{Optimal data size}

\paragraph{Channel learning example (N=4)}

We first consider learning of quantum channels with dimension $N = 4$, corresponding to 2 qubits, and assume that the Kraus rank is 2. We evaluate the performance of the four methods discussed earlier. For ALS-$N^2$, we compare two scenarios: (1) using $M$ \textbf{random} initial states and random observables, and (2) using \textbf{fixed} first-row initial states with $M_O$ random observables. We let $M = (3N-2)M_O = 10M_O$ to match the total number of measurements. We also compare the methods for choosing initial states and observables, either uniformly iid Pauli matrices or random methods using \texttt{rand\_dm} and \texttt{rand\_herm}. The recovery rate is defined as the number of times the Frobenius error is less than $10^{-5}$, over 20 total tests. 

We used the same numerical setting as in \cite[Figure 3]{klieschGuaranteedRecoveryQuantum2019} to compare with those methods. In \cite{baldwinQuantumProcessTomography2014}, the authors introduced the CPT-fit method, which has been proven for its efficiency and numerically compared with other gradient descent-based methods in \cite{klieschGuaranteedRecoveryQuantum2019}. We note that the gradient descent-based methods need computation time ranges from 10 to 140 seconds.

\begin{figure}[t!]
    \centering
    \begin{subfigure}[t]{0.5\textwidth}
        \centering
        \includegraphics[width = \textwidth]{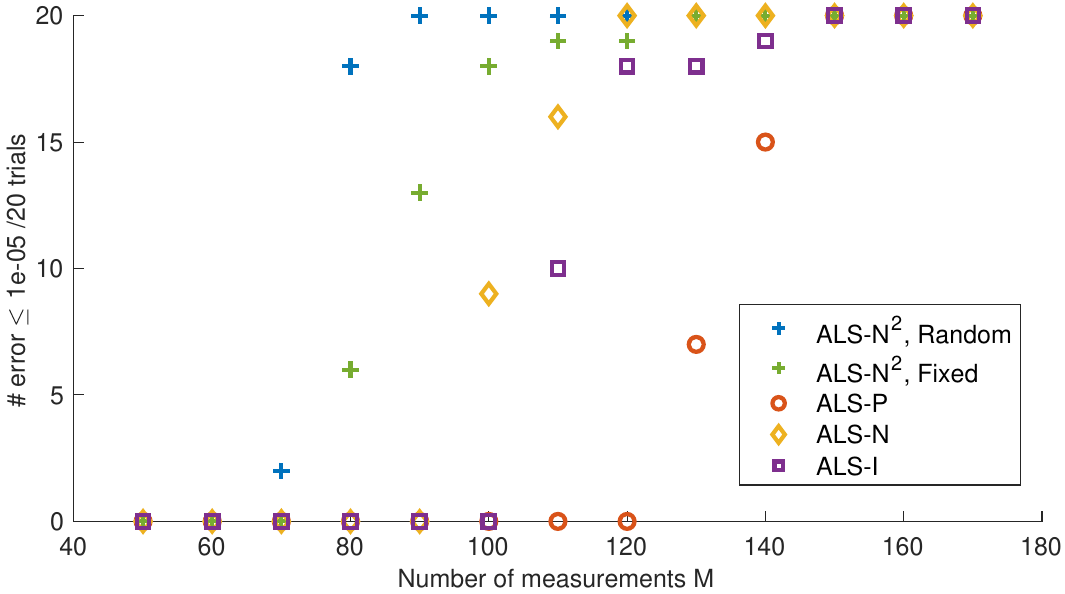}
        \caption{Random observables and density matrices.}
    \end{subfigure}%
    ~ 
    \begin{subfigure}[t]{0.5\textwidth}
        \centering
        \includegraphics[width = \textwidth]{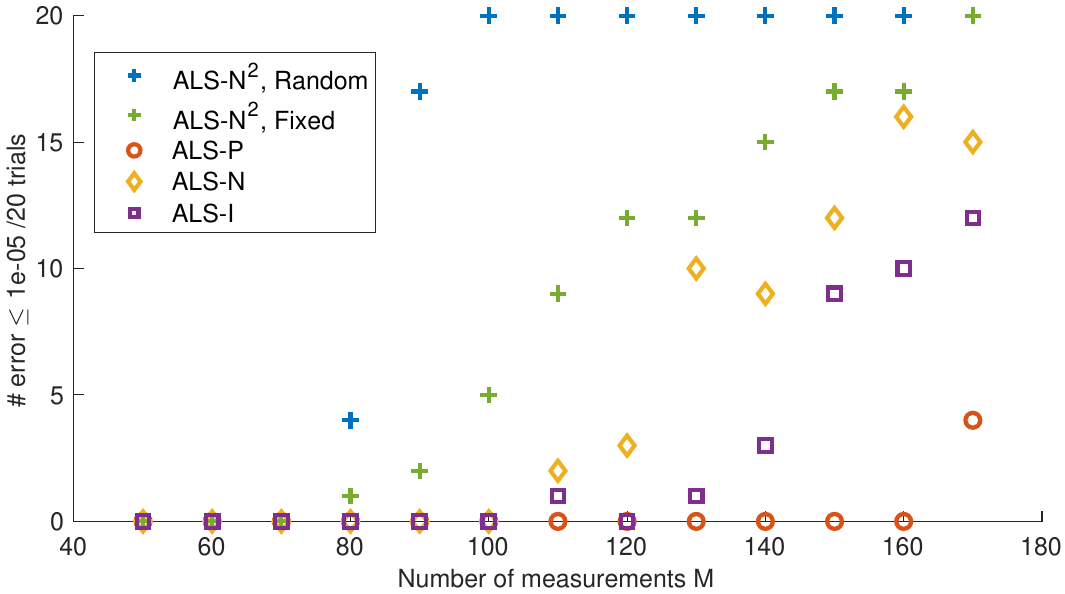}
        \caption{Pauli observables and density matrices.}
    \end{subfigure}
    
    \caption{
    The recovery rate for different numbers of measurements in channel learning with $N=4$. The ALS-$N^2$ using random density matrices and observables achieves the highest recovery rate, comparable to the CPT-fit method in \cite{klieschGuaranteedRecoveryQuantum2019}. Using ALS-$N^2$ with fixed initial states has slightly lower recovery rates due to decreased coherence in the joint tensor product $\overline{\rho}_0\otimes O$ when $\rho_0$ is fixed. For first-row blockwise measurements, ALS-N outperforms ALS-I. ALS-P requires more measurements to achieve successful recovery because of the lack of synchronization in the column space. The required number of measurements differs at most a constant multiple, which is independent of the rank $r$ and the system size $N$. Using Pauli matrices for observables and initial states requires more measurements for all methods. Although Pauli measurements offer performance guarantees, we argue that the random design achieves better performance due to the redundancy introduced by uniform iid sampling in the selection of Pauli matrices.
}
\label{fig_Sec_5_2_rate}
\end{figure}

The recovery rate is presented in Figure \ref{fig_Sec_5_2_rate}. The previous example demonstrated the time efficiency of the first-row recovery, and in this experiment, all methods finished the tasks within 0.2 seconds, which is at least 50 times faster than gradient descent-based methods. Since the scaling of computation time is not significant, we postponed the study of computation time in the next example. 



\begin{figure}[t!]
    \centering
    \begin{subfigure}[t]{0.5\textwidth}
        \centering
        \includegraphics[width = \textwidth]{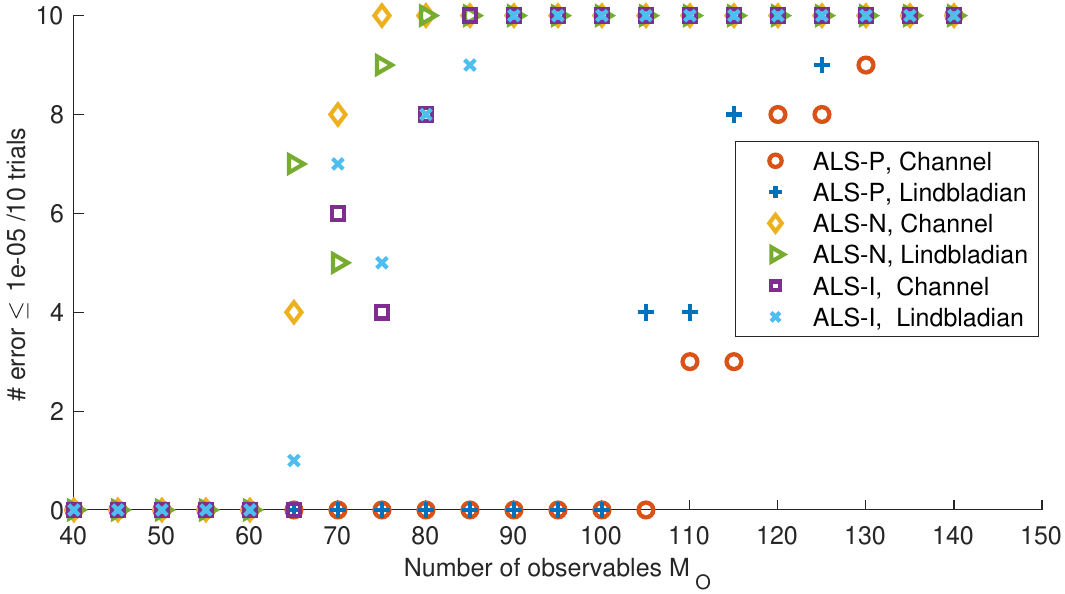}
        \caption{Recovery rate.}
    \end{subfigure}%
    ~ 
    \begin{subfigure}[t]{0.5\textwidth}
        \centering
        \includegraphics[width = \textwidth]{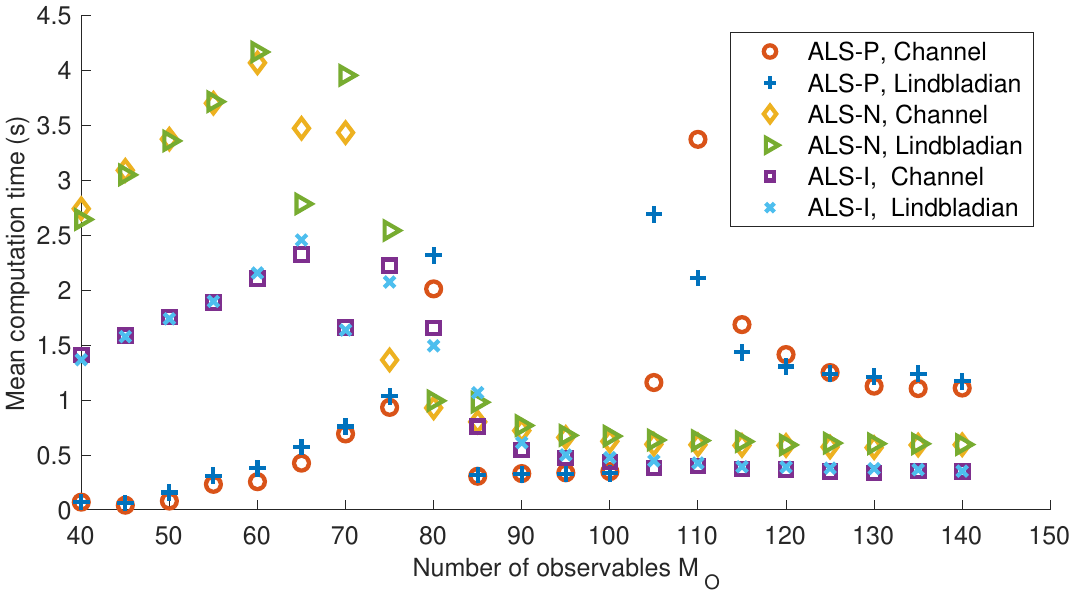}
        \caption{Computation time.}
    \end{subfigure}
    
    \caption{
    The recovery rate and computation time for channel and Lindbladian learning with $N = 16$. The channels and the Lindbladians have the reshaped rank of 3. Their performance is roughly the same. 
    ALS-I requires slightly more data and is faster than ALS-N. Initially, the computation time for all methods grows linearly with $M_O$ since methods fail to converge within a preset number of iterations. After reaching the successful recovery threshold for $M_O$, all algorithms provide correct solutions with fewer iterations, and computation time stabilizes. The ALS-P has a valley in computation time that precedes correct recovery since the algorithm converges for submatrices. However, the lack of synchronization prevents accurate recovery of the full matrix, which explains why ALS-P needs more data. 
}
    \label{fig_Sec_5_2_large}
\end{figure}

\paragraph{Channel and Lindbladian learning example (N=16)}
We did a similar test in a larger system size with $N = 16$, hence $n = 4$ qubits, and assumed rank $r = 3$, which corresponds to the middle system size in Section \ref{sec_representative_example}. We evaluate the recovery rate and computation time for both channel and Lindbladian learning using only fixed blockwise measurement design with methods ALS-P, ALS-N, and ALS-I. The Lindbladians are assumed to have 1 jump operator, resulting in a reshaped matrix $\bK$ of rank 3. As shown in Table \ref{table_Lindbladian}, the computation time for the random measurement design with ALS-$N^2$ is approximately 476 seconds, while the blockwise measurement design requires at most 2 seconds. Therefore we only focus on the results of the blockwise measurement design. Additionally, we use random observables, as they outperform Pauli observables in this context. The results are presented in Figure \ref{fig_Sec_5_2_large}. The channel and Lindbladian learning produce comparable outcomes and demonstrate the effectiveness of our methods for both tasks.

\subsection{Convergence with number of observables $M_O$.}\label{sec_noisy_M}
\begin{figure}[t!]
    \centering
    \includegraphics[width = \textwidth]{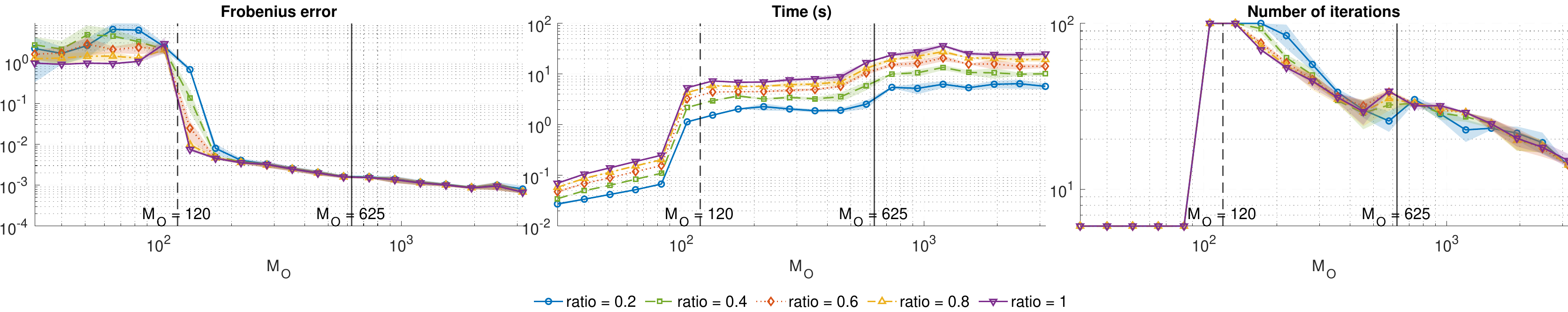}
    \caption{The change of Frobenius error, computation time, and number of iterations in ALS, with the number of observables $M_O$, in a noisy recovery of Lindbladian with 2 jump operators. The Frobenius error decays to less than $10^{-2}$ after $M_O = 120$, indicating a successful recovery. The error decays further with $M_O$ at an order of ${-0.68}$, fitted by a straight line, which agrees with the theoretical guarantee. After a successful recovery, we empirically observe the computation time increases with an order of $M_O^{0.5}$, and the number of iterations decreases with an order of $M_O^{-0.5}$. The necessary data size $M_O$ enlarges if we reduce the sampling ratio, which also reduces the computation time accordingly. }
    \label{fig_Sec_5_3_noisy_M}
\end{figure}

In this work, we investigate the convergence of the Frobenius error under observation noise as the number of observables $M_O$ increases, together with the change of computation time and the number of ALS iterations, in the setting of Lindbladian learning. We consider a Hilbert space of dimension $N=25$, where the Lindbladians have 2 jump operators and hence $\rank(\bK^\star) = 4$.  For $M_O$ observables, there are $M = 73M_O$ measurements, and the assumed noise level is $\sigma = 10^{-3}$. We compare the ALS-I method using subset sampling ratios of $0.2$, $0.4$, $0.6$, $0.8$, and $1$, whereas the last case corresponds to the ALS-N method. For tomography of a single block, the minimal number of observables $M_O$ required is on the order of $rN = 100$, while there are $N^2 = 625$ unknowns in that block. We mark in the figure with $M_O$ = 120 as sufficient for a successful recovery and $M_O$ = 625 as a full tomography. We extend $M_O $ beyond $N^2$ to further explore the overall convergence behavior. The result is presented in Figure \ref{fig_Sec_5_3_noisy_M}. For each case, we report the mean of 10 independent samples, and the shaded range indicates one standard deviation from the mean.

\subsection{Necessary number of observables with changing $N$ and $r$}
\begin{figure}[t!]
    \centering
    \begin{subfigure}[t]{0.5\textwidth}
        \centering
        \includegraphics[width = \textwidth]{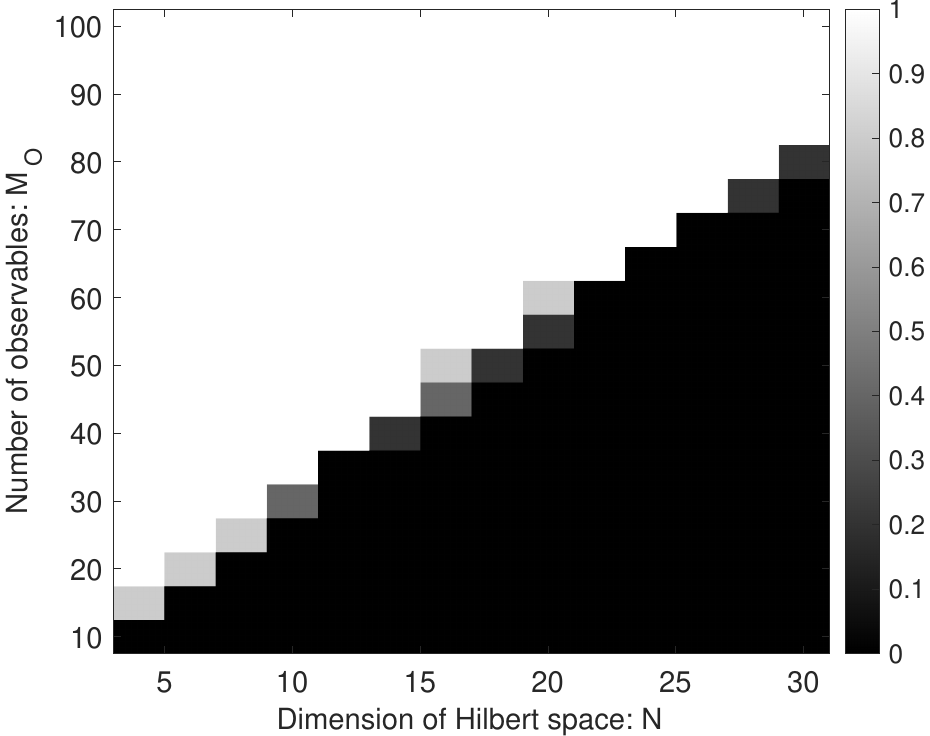}
        \caption{Channel recovery rate, fixed $r=2$, changing $N$.}
        \label{fig_sec_5_4_1}
    \end{subfigure}%
    ~ 
    \begin{subfigure}[t]{0.5\textwidth}
        \centering
        \includegraphics[width = \textwidth]{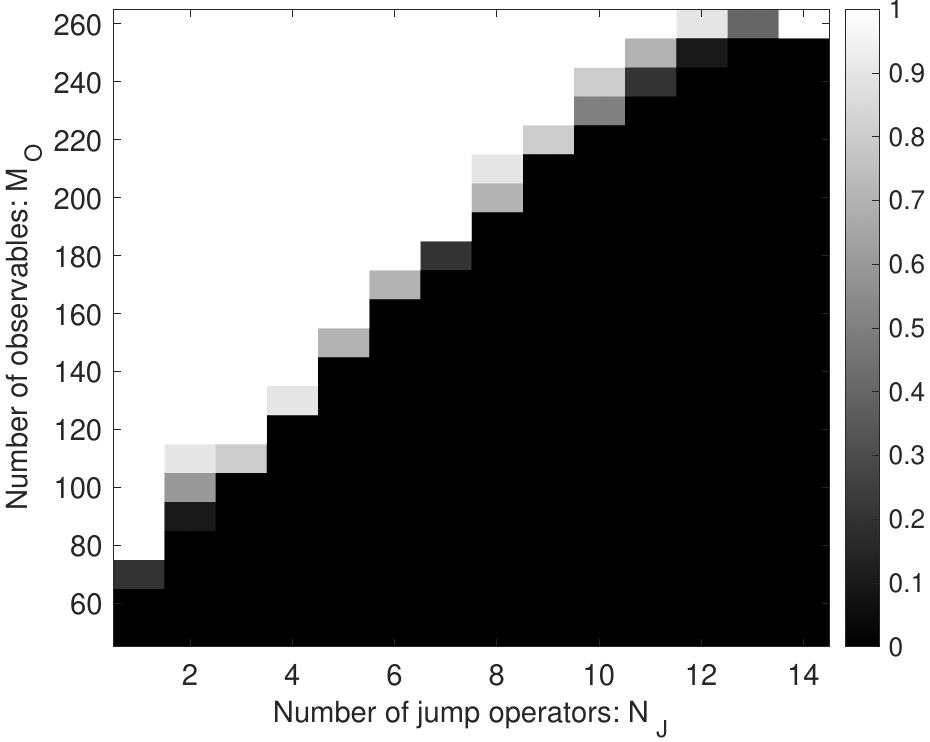}
        \caption{Lindbladian recovery rate, fixed $N=16$, changing $N_J$.}
        \label{fig_sec_5_4_2}
    \end{subfigure}
    
    \caption{Recovery rate with changing $N$ and $r$. The growth of $M_O$ is roughly linear in $N$ and $r$, so the number of measurements $M$ scales roughly quadratically in $N$ and linearly in $r$. 
}
    \label{fig_Sec_5_4_large}
\end{figure}

We compare the optimal data size $M_O$ with different dimensions of Hilbert space $N$ and rank $r$. In the first test, we learn the quantum channels with Kraus rank 2. We use ALS-I with a subset ratio of 0.5 for different values of $N$ and $M$ and repeat the test 10 times. We mark a successful recovery if the relative Frobenius error is less than $10^{-5}$, with no observation noise. The result is presented in Figure \ref{fig_sec_5_4_1}. We shall see that the number of $M_O$ required grows linearly with $N$. Hence, the number of measurements $M$ grows quadratically with $N$,  up to a difference in the power of $\log N$. 

In the second test, we learn the Lindbladian with fixed Hilbert space of dimension $N = 16$, where we change the number of jump operators from 1 to 14. Note that when $N_J = 14$, the rank of $\bK^\star$ equals to $N = 16$. We repeat the test 10 times for each parameter setting and record the recovery rate as before. The result is presented in Figure \ref{fig_sec_5_4_2}. When $r$ is small, $M_O$ grows linearly with $r$ until $r$ reaches $N$, which stops at the level of $N^2$ since there are in total $N^2$ unknowns in a single block. Note that the recovery rate when $M_O = 260, N_J = 12, 13$, is not 1. Recall the problem of learning the first block, we reparametrized the problem into learning $\bK_{11}= UV$ and actually introduced more variables when $r \geq N/2$. This redundancy is noticeable when $r$ reaches the size of $N$, hence needs more data than necessary for a successful recovery. Here we focus on a low-rank recovery scenario, whereas $r \leq N$. See the related discussion in Section \ref{sec_3_r_geq_N} to extend our method when $N \leq r \leq N^2$.

\section{Conclusion}\label{sec_conclusion}
This work introduces a unified and efficient approach for quantum superoperator learning, addressing both quantum channel and Lindbladian reconstruction via low-rank matrix sensing. By leveraging the restricted isometry property (RIP) of measurement operators, we provide theoretical guarantees for the identifiability and recovery of low-rank superoperators. Our methodology, including alternating least squares with acceleration, efficiently exploits the operators' low-rank structure and ensures scalability as the system dimension grows.

Numerical experiments validate the proposed approach, demonstrating robust performance in noisy cases. The flexibility of our approach to accommodate both positive semidefinite and non-positive semidefinite superoperators highlights its versatility in quantum information processing.

Future directions include extending to incorporate noise models reflective of experimental conditions and exploring adaptive measurement designs to further reduce the number of required observations. These advancements could significantly enhance the practical utility of quantum superoperator learning in experimental and theoretical research.

\section*{Acknowledgment}
We thank Bowen Li for the helpful discussions. The work is supported in part by the National Science Foundation through the DMS-2309378 and IIS-2403276 awards.



\printbibliography
\end{document}